\documentclass[journal]{IEEEtran}

\usepackage[table,xcdraw]{xcolor}  
\usepackage{colortbl,hhline}              

\usepackage{amsmath,amssymb,amsfonts}

\usepackage{graphicx}
\graphicspath{{../pdf/}{../jpeg/}}
\DeclareGraphicsExtensions{.pdf,.jpeg,.png}

\usepackage[caption=false,font=footnotesize]{subfig}

\usepackage{booktabs}
\usepackage{multirow}
\usepackage{array}
\usepackage{diagbox}

\usepackage{algorithm}
\usepackage{algpseudocode}

\usepackage{pifont}

\usepackage{tikz}
\usetikzlibrary{arrows.meta, positioning, shapes, calc}

\usepackage{cite}
\usepackage{url}
\usepackage{verbatim}
\usepackage{stfloats}
\usepackage{balance}
\usepackage{bbm}
\usepackage{arydshln}
\usepackage{enumitem}

\usepackage{hyperref}

\begin{document}
\title{$\mathcal{U}^3$-xi: Pushing the Boundaries of Speaker Recognition by Incorporating Uncertainty}
\author{Junjie Li, Kong Aik Lee$^\dagger$,~\IEEEmembership{Senior Member,~IEEE}

\thanks{Junjie Li and Kong Aik Lee are with the Department of Electrical and Eletronic Engineering, Faculty of Engineering, and Research Centre for Data Science and Artificial Intelligence (RC-DSAI), The Hong Kong Polytechnic University, Hong Kong SAR, China (email: junjie98.li@connect.polyu.hk, kong-aik.lee@polyu.edu.hk ). \\
$\dagger$: Corresponding author
}}
    
\maketitle

\begin{abstract}
An utterance-level speaker embedding is typically obtained by aggregating a sequence of frame-level representations. However, in real-world scenarios, individual frames encode not only speaker-relevant information but also various nuisance factors.  As a result, different frames contribute unequally to the final utterance-level speaker representation for Automatic Speaker Verification.
To address this issue, we propose to  estimate the inherent uncertainty of each frame and assign adaptive weights accordingly, where frames with higher uncertainty receive lower attention. Based on this idea, we present $\mathcal{U}^3$-xi (pronounced as `u-cube-xi'), 
a comprehensive framework designed to produce more reliable and interpretable 
uncertainty estimates for speaker embeddings.
Specifically, we introduce several strategies for uncertainty supervision. First, we propose \textit{speaker-level uncertainty supervision} via a \textit{Stochastic Variance Loss}, where the stochastic estimate of the embedding deviation serves as a pseudo ground truth for uncertainty learning. 
Second, we incorporate \textit{global-level uncertainty supervision} by injecting the predicted uncertainty into the $softmax$ scale during training. This adaptive scaling mechanism adjusts the sharpness of the decision boundary according to sample difficulty, providing global guidance. Third, we redesign the uncertainty estimation module by integrating a Transformer encoder with \textit{multi-view self-attention}, enabling the model to capture rich local and long-range temporal dependencies.
Comprehensive experiments demonstrate that $\mathcal{U}^3$-xi is model-agnostic and can be seamlessly applied to various speaker encoders. In particular, when applied to ECAPA-TDNN-512, it achieves 21.1\% and 15.57\% relative improvements on the VoxCeleb1 sets in terms of EER and minDCF, respectively\footnote{The code can be found in: \textcolor{magenta}{\url{https://github.com/mrjunjieli/wespeaker_u_cube.git}}}. 

\end{abstract}


\section{Introduction}

\IEEEPARstart{A}{utomatic} speaker verification (ASV) aims to determine a speaker’s identity solely from their voice using machine learning algorithms and is one of the most convenient and natural forms of biometric authentication \cite{singh2018voice}. ASV systems are designed to extract speaker-discriminative information while suppressing nuisance factors such as linguistic content, emotional variation, and background noise \cite{wang2024overview}.
Beyond biometric authentication, ASV supports a wide range of applications, including personalized user services \cite{mcgraw2016personalized} as well as security and surveillance scenarios \cite{kiktova2015speaker}. Moreover, speaker modeling techniques developed for ASV are fundamental to various downstream tasks, such as target speaker extraction \cite{li2024effectiveness}, speech synthesis \cite{du2024cosyvoice}, and voice conversion \cite{lu2019one}, where high-quality speaker embeddings are indispensable

A typical speaker recognition system consists of three modules: a front-end speaker encoder, a pooling layer, and a classifier. The speaker encoder transforms the input speech waveform into a sequence of frame-level embeddings. However, speech is inherently complex and contains not only speaker traits but also numerous non-speaker variations \cite{wang2024overview}. To summarize the temporal sequence into a fixed-dimensional representation, the pooling layer aggregates frame embeddings with weights assigned adaptively to different frames. Although many pooling strategies have been proposed \cite{li2017deep,snyder2018x,wang2021revisiting, okabe2018attentive, zhu2018self,india2019self, zhao2022multi, wu2020vector,desplanques20_interspeech, ma2025expo}, most approaches still overlook data uncertainty, which has been shown to be beneficial across multiple domains \cite{chang2020data,ou2021sdd,shi2019probabilistic,ji2023map,meng2021magface,chen2022fast,lee2021xi,reynolds2000speaker,dehak2010front,liu2023disentangling,wang24ha_interspeech,chen2024modeling,chen2024pseudo}.


Recently, Lee et al. \cite{lee2021xi} introduced uncertainty estimation into deep neural networks for ASV through the xi-vector model. This approach applies Gaussian posterior inference to aggregate temporal frame-level features into an utterance-level speaker embedding while simultaneously estimating uncertainty. 
Despite its success, the xi-vector framework still exhibits several limitations that restrict the reliability of its uncertainty estimation.
First, the uncertainty estimation module consists of only two shallow linear layers, which may lack sufficient modeling capacity to capture complex non-linear relationships across temporal frames. Second, the xi-vector is trained solely with a cross-entropy (CE) loss, for which the uncertainty is learned implicitly through classification supervision alone, without any explicit guidance or direct supervision on the uncertainty estimate itself. This leads to sub-optimal uncertainty modeling and reduces robustness in real-world scenarios.

To address the above limitations, we propose a series of improvements aimed at producing more reliable and interpretable uncertainty estimation:
\begin{itemize}
    \item  \textbf{Speaker-level Uncertainty Supervision.}
We use stochastic estimate of embedding deviation
as a pseudo–ground truth target to explicitly supervise the predicted uncertainty. This forms our \textit{Stochastic Variance Loss} (SVL). Furthermore, the predicted uncertainty is incorporated into the cosine scoring function, leading to an \textit{uncertainty-aware cosine scoring} scheme. 
\item \textbf{Global-level Uncertainty Supervision.}
Speaker-level supervision provides guidance only among utterances belonging to the same speaker, but it fails to capature uncertainty across different speakers. To address this limitation, we introduce global-level uncertainty supervision by introducing uncertainty into the scale parameter of the $softmax$ function. The scale dynamically adapts to sample difficulty, making the $softmax$ distribution sharper for easy samples and smoother for hard ones. Through this mechanism, uncertainty is explicitly learned at the global distribution level, resulting in a 3.45\% relative improvement. 

\item \textbf{Uncertainty Estimation Module.}
To enhance the modeling capacity of uncertainty estimation, we incorporate a Transformer encoder \cite{vaswani2017attention} into the uncertainty estimation module. The self-attention mechanism significantly improves the network’s ability to model complex non-linear relationships and capture long-range temporal dependencies—capabilities that shallow linear layers inherently lack. Furthermore, to better encode short to long range temporal structures, we extend standard multi-head self-attention (MHA) to a \textit{multi-view self-attention} (MVA) \cite{wang2022multi} design. This extension enables richer and more robust variance estimation across both short- and long-range temporal dependencies, resulting in a 5.48\% relative improvement.
\end{itemize}

Each of the proposed enhancements contributes to performance improvement. When combined, they yield a substantial cumulative gain, achieving 21.1\% and 15.57\% relative improvements in terms of equal error rate (EER) and minium detection cost (minDCF), respectively, compared with the  original ECAPA-TDNN model.
\section{Related work}

\subsection{Uncertainty Estimation in Deep Learning}

In general, uncertainty can be categorized into two types: epistemic and aleatoric \cite{der2009aleatory,mukhoti2021deterministic,mukhoti2023deep}. \textit{Epistemic uncertainty} arises from the model or its parameters, representing the model’s lack of knowledge. For instance, when the test utterances lie outside the training distribution, the model should exhibit high epistemic uncertainty. In contrast, \textit{aleatoric uncertainty} originates from the data itself, reflecting inherent noise or ambiguity. A higher aleatoric uncertainty indicates that the data contain greater noise or variability. Similar to \cite{lee2021xi}, this paper focuses exclusively on the uncertainty intrinsic to the data. Therefore, unless otherwise specified, the term uncertainty refers to \textit{aleatoric uncertainty}.

In real-world scenarios, data are rarely perfectly clean. They may contain irrelevant information, background noise, or even incorrect labels, all of which contribute to varying degrees of data uncertainty. Such uncertainty arises from multiple sources and can manifest differently across modalities, such as speech and visual data. Effectively modeling this uncertainty is essential for improving the robustness and interpretability of deep neural networks (DNNs) \cite{chang2020data,ou2021sdd,shi2019probabilistic,ji2023map,meng2021magface,chen2022fast}. These studies introduced Gaussian representations in the latent space encoded by DNNs, where the variance serves as an estimate of the uncertainty. 

In speaker recognition task, the xi-vector model \cite{lee2021xi} was proposed to estimate the uncertainty of each speech frame using deep neural networks (DNNs), inspired by the successful treatment of uncertainty in the \textit{universal background model} (UBM) \cite{reynolds2000speaker} and the i-vector framework \cite{dehak2010front}. Similar concepts have  been widely adopted and shown effective in related areas such as speech disentanglement \cite{liu2023disentangling}, voice anonymization \cite{wang24ha_interspeech,chen2024modeling,chen2024pseudo}, and speaker verification \cite{li2025xi+, wang2023incorporating, wang2024cosine,barahona2025analysis}. 




\subsection{Large Margin Softmax Loss Function}

Large margin softmax loss functions have been extensively studied and shown to be
highly effective in both face recognition
\cite{liu2016large,liu2017sphereface,wang2018cosface,deng2019arcface,wang2020mis,
huang2020curricularface,kim2022adaface}
and speaker verification tasks
\cite{zhou2020dynamic,li2018angular,liu2019large,li2022real}.
By introducing an angular or additive margin to the target class, these losses
enforce stricter class separation and encourage intra-class compactness.
Early works such as SphereFace \cite{liu2017sphereface} introduced multiplicative
angular margins, while subsequent methods, including CosFace (additive cosine
margin) \cite{wang2018cosface} and ArcFace (additive angular margin)
\cite{deng2019arcface}, further improved the geometric interpretability and
optimization stability.
In the speaker verification domain, large margin softmax variants (e.g.,
AAM-Softmax \cite{deng2019arcface}, AM-Softmax \cite{wang2018cosface}) have become
standard training objectives due to their strong discriminative capability.

In addition to margin design, several studies
\cite{liu2025adaspeaker,shang2023improving,xuan2025exploring,zhang2019adacos,
zhu2023coarse,kail2023scaleface}
have investigated the role of the scale parameter, which controls the sharpness
of the posterior distribution produced by the softmax layer.
A larger scale amplifies logit differences, yielding sharper posteriors and
stronger class separation, but may lead to over-confidence when embeddings are
unreliable.
Conversely, a smaller scale produces smoother posteriors and more stable
gradients, particularly under challenging acoustic conditions, at the cost of
reduced discriminability.
Different works exploit this trade-off in various ways.
For example, a smaller scale has been shown to improve adversarial robustness
\cite{xuan2025exploring}, while AdaCos \cite{zhang2019adacos} adopts a dynamic scale
schedule.
In contrast, other methods increase the scale during training to emphasize hard
samples \cite{zhu2023coarse,liu2025adaspeaker}.
Beyond manually designed strategies, several approaches predict sample-wise
scales using DNNs \cite{shang2023improving,kail2023scaleface}, where the learned
scale can be interpreted as a \textit{quality} or \textit{uncertainty} indicator.

\section{Background}
\subsection{Uncertainty-aware Speaker Modeling with xi-vector}

In speaker  recognition tasks, generating  compact and discriminative embeddings is essential for representing speaker characteristics across speech utterances of varying length. The xi-vector framework \cite{lee2021xi, wang2023incorporating,liu2023disentangling, wang2024cosine} introduces a novel temporal pooling mechanism based on Gaussian posterior inference, which aggregates frame-level features into a fixed-dimensional utterance embedding while weighting each frame according to its estimated uncertainty.

\begin{figure}[tbp]
    \centering
    \includegraphics[width=0.9\linewidth]{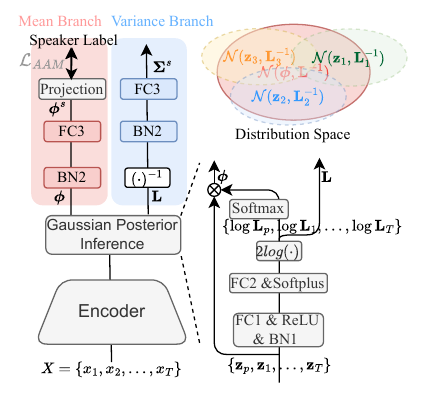}
    \vspace{-3mm}
\caption{Architecture of the xi-vector model with integrated uncertainty estimation.
The mean and variance branches share the same parameters but apply different operations, reflecting the distinct mathematical formulations of mean and variance from a statistical perspective. }
\vspace{-3mm}
    \label{fig:xi}
\end{figure}

As illustrated in Fig.~\ref{fig:xi}, the Gaussian posterior inference serves as a generalizable pooling module that can be integrated into any speaker encoder architecture. Given an input speech sequence $X=\{x_1, x_2, \ldots, x_T\}$ of  $T$ frames, a speaker encoder first transforms it into a sequence of frame-level representations $\{\mathbf{z}_1, \mathbf{z}_2, \ldots, \mathbf{z}_T\}$. In addition, the model estimates the uncertainty associated with each frame, expressed as the log-precision  $\log \mathbf{L}_t$, where higher precision denotes lower uncertainty. To maintain computational efficiency, the precision matrix $\mathbf{L}_t$ is assumed to be \textbf{diagonal} in practice. 
From the perspective of a linear Gaussian model \cite{roweis1999unifying}, the observed representation $\mathbf{z}_t$ can be decomposed into two components:
\begin{equation}
\mathbf{z}_t = \mathbf{h} + \boldsymbol{\epsilon}_t,
\end{equation}
where $\mathbf{h} \sim \mathcal{N}(\boldsymbol{\mu}_p, \mathbf{L}_p^{-1})$ \footnote{The quantities $\boldsymbol{\mu}_p$ and $\mathbf{L}_p^{-1}$ denote the prior mean and prior covariance matrix, respectively. They are initialized as $0$ and $1$. For notational consistency with the frame-level representations $\mathbf{z}_t$, we also denote the prior mean $\boldsymbol{\mu}_p$ as $\mathbf{z}_p$ in the subsequent sections. Both symbols refer to the same quantity.} denotes the latent speaker variable shared across the entire sequence, and $\boldsymbol{\epsilon}_t \sim \mathcal{N}(0, \mathbf{L}_t^{-1})$ represents the frame-specific random variable capturing uncertainty at time $t$, respectively. 
Given current observation sequence, the posterior distribution of $\mathbf{h}$ is estimated as follow (see \cite{lee2021xi} for derivation): 
\begin{equation}
p(\mathbf{h}|\mathbf{z}_1,...,\mathbf{z}_T,\mathbf{L}_1^{-1},...,\mathbf{L}_T^{-1})=\mathcal{N}(\mathbf{h}|\boldsymbol{\phi},\mathbf{L}^{-1}). 
\end{equation}
By considering its prior mean and covariance, the posterior mean and covariance could be computed as: 
\begin{align}
    \boldsymbol{\phi} &= \frac{\sum_{t=1}^{T}\mathbf{L}_t\mathbf{z}_t+\mathbf{L}_p\mathbf{z}_p}{\mathbf{L}} \text{ and} \\
    \mathbf{L}^{-1} &= \left( \sum_{t=1}^T\mathbf{L}_t + \mathbf{L}_p \right)^{-1}. 
\end{align} 
Then, the posterior mean $\mathbf{\phi}$ and covariance  $\mathbf{L}^{-1}$ are propagated through a batch normalization layer followed by a fully connected layer to obtain the final speaker embedding. Following the property of Gaussian distribution $\mathcal{N}(\mathbf{h}|\boldsymbol{\phi},\mathbf{L}^{-1})$, where the transform on $\mathbf{h}$  is applied conserving its mean and covariance matrix. This is implemented as shown in Fig. \ref{fig:xi}.   The parameters of these layers are shared between the mean branch and the variance branch, but different statistical operations are applied in each branch \cite{wang2023incorporating,chen2024pseudo}.

For the mean branch, the transformation can be expressed as\footnote{These equations are written in PyTorch style for clarity, similar to \url{https://docs.pytorch.org/docs/stable/generated/torch.nn.BatchNorm1d.html}.}
:
\begin{equation}
\boldsymbol{\phi}^\text{s} = \left(\frac{\boldsymbol{\phi} - \boldsymbol{\mu}_\text{bn}}{\sqrt{\boldsymbol{\sigma}_\text{bn} + \epsilon \mathbf{I}} } \otimes \boldsymbol{\gamma}_\text{bn} + \boldsymbol{\beta}_\text{bn}\right) \mathbf{A}_\text{fc}^\top + \mathbf{b}_\text{fc},
\end{equation}
where $\boldsymbol{\mu}_\text{bn}$ and $\boldsymbol{\sigma}_\text{bn}$ denote the batch-wise mean and variance in batch normalization, respectively. The parameter $\epsilon$ is a small constant for numerical stability, and $\boldsymbol{\gamma}_\text{bn}$ and $\boldsymbol{\beta}_\text{bn}$ are learnable affine parameters. The symbol $\otimes$ denotes element-wise multiplication. Finally, $\mathbf{A}_\text{fc}$ and $\mathbf{b}_\text{fc}$ are the learnable parameters of the fully connected layer.

For the variance branch, the same set of parameters as in the mean branch is used, but with a different computational formulation in order to model the covariance:
\begin{equation}
\mathbf{\Sigma}^\text{s} = \mathbf{A}_\text{fc}  \frac{\mathbf{L}^{-1} \otimes \boldsymbol{\gamma}_\text{bn}^2}{\boldsymbol{\sigma}_\text{bn} + \epsilon\mathbf{I}} \mathbf{A}_\text{fc}^\top. 
\end{equation}

\subsection{AAM-Softmax for Speaker Classification}

During training, the utterance-level representation
$\boldsymbol{\phi}^\text{s}$ is projected into a $C$-dimensional space,
where $C$ denotes the number of speakers in the training set.
Speaker classification is performed using Additive Angular Margin Softmax
(AAM-Softmax) \cite{deng2019arcface}, which is widely adopted in speaker
recognition to enhance embedding discriminability.

AAM-Softmax removes the bias term and applies $\ell_2$ normalization to both
the speaker embeddings and the classifier weights, such that classification
depends solely on their cosine similarity.
An additive angular margin $m$ is imposed on the target class to explicitly
encourage intra-class compactness and inter-class separability.
The resulting loss function is defined as:
\begin{equation}
\mathcal{L}_\text{AAM}
= -\frac{1}{B} \sum_{i=1}^B
\log
\frac{e^{s \cos (\theta_{y_i} + m)}}
{e^{s \cos (\theta_{y_i} + m)} + \sum_{j=1,j \neq y_i}^{C} e^{s \cos \theta_j}},
\label{eq:AAM}
\end{equation}
where $B$ is the number of training samples in a batch, and $y_i$ denotes the ground-truth
speaker label of the $i$-th utterance.
The quantity $\theta_j$ represents the angle between the normalized embedding
$\boldsymbol{\phi}_i^\text{s}$ and the $j$-th normalized class weight.
The scaling factor $s$ controls the magnitude of the logits, while $m$
determines the enforced angular margin for the target class.

\section{Speaker-level Uncertainty Supervision }

\subsection{Stochastic Variance Loss}

Our previous work on xi-vector  \cite{lee2021xi} relies solely on a classification loss to \textit{implicitly} supervise frame-level uncertainty, without any form of explicit supervision. Such implicit uncertainty learning often results in suboptimal or unstable  estimation. To overcome this limitation, we introduce an \textit{explicit} supervision mechanism for utterance-level uncertainty modeling, termed \textit{Stochastic Variance Loss} (SVL). Our initial work \cite{li2025xi+} firstly proposed SVL in the context of speaker recognition and demonstrated its effectiveness. In this paper, we further generalize this idea by exploring other alternative formulations to realize the core concept of SVL.

The proposed loss $\mathcal{L}_\text{SVL}$ aims to establish a pseudo ground truth for utterance-level uncertainty estimation. Specifically, we treat the distance between the centroid of embeddings from the same speaker and the current utterance’s embedding $\boldsymbol{\phi}^\text{s}$ as a pseudo label. 

Suppose we use $\boldsymbol{\phi}^\text{c}_{y_i}$ to represent the centroid of speaker $y_i$ of utterance $i$: 
\begin{equation}
    \boldsymbol{\phi}^\text{c}_{y_i} = \frac{\sum_{i=1}^{I}  \boldsymbol{\phi}^\text{s}_i}{I},
\end{equation} where all $I$ utterances belong to the same speaker $y_i$. The loss $\mathcal{L}_\text{SVL}$ is given by: 
\begin{equation}
    \mathcal{L}_\text{SVL} = \frac{1}{B}  \sum_{b=1}^{B} \|\alpha \mathbf{(\Sigma}^{\text{s}}_b)^{\frac{1}{2}} -|\boldsymbol{\phi}^\text{s}_b-\boldsymbol{\phi}^\text{c}_{y_b}| \|^2,
    \label{equ:new_svl}
\end{equation}
where $B$ denotes the number of batch size in training. The matrices of current batch $\mathbf{\Sigma}^\text{s}_b$ and $\boldsymbol{\phi}^\text{c}_{y_b}$
denote the covariance matrix  for utterance $b$ and the centroid of speaker $y_b$, respectively. The symbol $\|\cdot\|$ denotes the Euclidean ($\ell_2$) norm.
 In addition, $\alpha$ is a learnable parameter initialized to one, referred to as the uncertainty scaling factor. It aimes to address potential scale mismatch between predicted uncertainty and pseudo ground truth. Since only the embedding distance of a single utterance-centroid pair is used to approximate the uncertainty (i.e., variance), we refer to this approach as \textit{stochastic variance} learning.

There are two  strategies for computing the centroids:
\begin{itemize}
\item \textbf{Fix:} Centroids are computed using embeddings $\boldsymbol{\phi}^\text{s}$ extracted from a pre-trained model with the same architecture, but was trained solely with the cross-entropy loss $\mathcal{L}_\text{AAM}$. As a result,  speaker centroids remain \textbf{fix}ed throughout training and are subsequently used to provide supervision, as defined in ~\eqref{equ:new_svl}. 

\item \textbf{Pro:} Instead of relying on a pre-trained model, the speaker embedding $\boldsymbol{\phi}^\text{s}$ is extracted using the checkpoint from the last epoch. Then centroids are computed  after each training epoch. This allows the centroids  to evolve along with model optimization, leading to \textbf{pro}gressively more accurate estimate.
\end{itemize}
The final training loss is given by: 
\begin{equation}
\mathcal{L}_{\text{Final}} = \mathcal{L}_{\text{AAM}}+ \kappa \mathcal{L}_\text{SVL},
\label{equ:loss}
\end{equation}
where $\kappa$  is used to control loss weight and  is dynamically adjusted based on the current epoch $\psi$ to  ensure that $\mathcal{L}_\text{SVL}$ 
  gradually contributes as training progresses: 
\begin{equation}
  \kappa = 
  \begin{cases}
    0, & \text{if } \text{epoch} \leq \psi_\text{SVL}, \\
   \lambda \frac{\psi-\psi_\text{SVL}}{\psi_\text{Max}-\psi_\text{SVL}}, & \text{otherwise},
  \end{cases}
\end{equation}
where $\lambda$ is a pre-defined fixed constant, $\psi_\text{SVL}$ indicates epoch when $\mathcal{L}_\text{SVL}$ starts to be applied. The hyperparameter $\psi_\text{Max}$ indicates the maximum number of training epochs.  In this paper, we set $\psi_\text{SVL}=40$    and $\psi_\text{Max}=150$. 


\subsection{Uncertainty-aware Scoring}
Since the uncertainty varies across utterances, it is intuitive to incorporate uncertainty into the scoring backend. In this work, we adopt cosine similarity as the scoring metric, which is computed as follows:
\begin{equation}
    sc_{u} = \frac{<\boldsymbol{\phi}^\text{s}_2,\boldsymbol{\phi}^\text{s}_1>}{\sqrt{\boldsymbol{(\phi}{^\text{s}_2})^\top (\mathbf{I}+\rho\boldsymbol{\Sigma}^{\text{s}}_2)^{-1}\boldsymbol{\phi}^\text{s}_2}  \sqrt{(\boldsymbol{{\phi}}{^{s}_1})^\top (\mathbf{I}+\rho\boldsymbol{\Sigma}^{\text{s}}_1)^{-1}\boldsymbol{\phi}^\text{s}_1}},
    \label{equ:cos}
\end{equation}
where $\rho$ denotes a scaling factor in cosine scoring, and $\mathbf{I}$ is the identity matrix.  When the covariance matrix $\boldsymbol{\Sigma}^s$ or $\rho$ is zero, the uncertainty-aware cosine scoring $sc_u$ reduces to the normal cosine similarity. Wang et al. \cite{wang2024cosine} have proposed to use the fixed constant $1/d$ as $\rho$, where $d$ denotes the dimension of speaker embedding.  However, a fixed value may not be optimal, as demonstrated in our previous work \cite{li2025xi+}. To address this limitation, we introduce a learnable uncertainty scaling factor  $\alpha$, which replaces $\rho$ in uncertainty-aware scoring. This parameter can be adapted to different training datasets, allowing the model to optimally calibrate the influence of uncertainty in the scoring process.

\section{Global-level Uncertainty Supervision}
Speaker-level uncertainty supervision can only compare uncertainty among utterance embeddings of the same speaker. For embeddings belonging to different speakers, the SVL loss does not provide a direct or meaningful basis for comparison. In this section, we introduce a global-level uncertainty supervision mechanism, which evaluates and constrains uncertainty across embeddings from all utterances in the training set, rather than being limited to within-speaker comparisons.

\subsection{Uncertainty-aware Softmax}

Conventional Softmax-based objectives, such as AAM-Softmax in (\ref{eq:AAM}), treat the speaker embedding $\boldsymbol{\phi}^\text{s}$ as a deterministic point estimate, ignoring its associated covariance $\mathbf{\Sigma}^\text{s}$. As a result, valuable information regarding embedding reliability and uncertainty is discarded.

To explicitly incorporate uncertainty, we reformulate the conventional $\ell_2$ normalization under an uncertainty-aware metric. The resulting normalized embedding is defined as:
\begin{align}
\tilde{\boldsymbol{\phi}}^\text{s}
= \frac{\boldsymbol{\phi}^\text{s}}{\|\boldsymbol{\phi}^\text{s}\|_{\mathbf{\Sigma}^\text{s}}}
&= \frac{\boldsymbol{\phi}^\text{s}}{\sqrt{(\boldsymbol{\phi}^\text{s})^\top (\mathbf{\Lambda} + \mathbf{\Sigma}^\text{s})\boldsymbol{\phi}^\text{s}}},
\label{equ:uncertainty}\\
&\propto {\boldsymbol{\phi}^\text{s}}'
\cdot
\frac{\sqrt{(\boldsymbol{\phi}^\text{s})^\top (\mathbf{\Lambda} + \mathbf{\Sigma}^\text{s})^{-1}\boldsymbol{\phi}^\text{s}}}{\|\boldsymbol{\phi}^\text{s}\|},
\end{align}
where ${\boldsymbol{\phi}^\text{s}}' = \boldsymbol{\phi}^\text{s} / \|\boldsymbol{\phi}^\text{s}\|$ denotes the conventional $\ell_2$-normalized embedding, and $\mathbf{\Lambda}$ is a positive-definite bias term introduced for numerical stability. This normalization constrains $\tilde{\boldsymbol{\phi}}^\text{s}$ to lie on the unit hypersphere under an uncertainty-aware (Mahalanobis) metric, thereby encoding both embedding direction and confidence.

In practice, explicitly computing the inverse $(\mathbf{\Lambda} + \mathbf{\Sigma}^\text{s})^{-1}$ can cause numerical instability and unreliable gradients. Therefore, we adopt the formulation in (\ref{equ:uncertainty}) as a stable surrogate during training. 
We further rewrite (\ref{equ:uncertainty}) as:
\begin{align}
\tilde{\boldsymbol{\phi}}^\text{s}
= {\boldsymbol{\phi}^\text{s}}' \cdot s_u ,
\end{align}
where
\begin{align}
s_u =
\frac{\|\boldsymbol{\phi}^\text{s}\|}{\sqrt{(\boldsymbol{\phi}^\text{s})^\top (\mathbf{\Lambda} + \mathbf{\Sigma}^\text{s})\boldsymbol{\phi}^\text{s}}}
\end{align}
is an \emph{uncertainty-aware scale} that compactly captures the effect of embedding uncertainty.
Notably, $s_u$ increases as uncertainty decreases. 

Finally, we incorporate the uncertainty-aware embedding into AAM-Softmax, resulting in the proposed uncertainty-aware AAM-Softmax (\textbf{UAAM}):
\begin{align}
\mathcal{L}_{\text{UAAM}}
= -\frac{1}{B} \sum_{i=1}^B
\log
\frac{
e^{s \cdot s_u \cos (\theta_{y_i} + m)}
}{
e^{s \cdot s_u \cos (\theta_{y_i} + m)}
+ \sum_{j=1, j\neq y_i}^{C} e^{s \cdot s_u \cos \theta_j}
}. 
\label{eq:UAAM}
\end{align} When $s_u = 1$, $\mathcal{L}_{\text{UAAM}}$ reduces to the conventional AAM-Softmax loss $\mathcal{L}_{\text{AAM}}$.

\subsection{Gradient-Driven Adaptation of Scale and Covariance}
From a training perspective, the uncertainty-aware scale $s_u$ enables the model
to adaptively modulate the loss according to the reliability of each embedding.
Since gradient-based optimization always aims to minimize the training loss,
the direction in which $s_u$ is adjusted depends on how loss reduction can be
most effectively achieved for different samples.

For unreliable or incorrectly classified samples, which typically incur large
loss values, reducing the scale $s_u$ provides a natural descent direction.
A smaller scale smooths the softmax distribution, attenuating the influence of
incorrect logits and thereby reducing the loss.
In our formulation, this decrease in $s_u$ directly corresponds to an increase
in the predicted variance (uncertainty), reflecting the higher ambiguity of hard
samples.

Conversely, for easy and reliable samples that are already correctly classified
and incur small losses, the optimizer still seeks further loss minimization.
In this regime, increasing $s_u$ sharpens the softmax distribution and amplifies
the target-class logit, pushing the predicted posterior probability closer to one.
Under the cross-entropy loss, this leads to additional loss reduction, and thus
the back-propagation gradient naturally drives $s_u$ upward.
As a result, the predicted variance $\mathbf{\Sigma}^\text{s}$ decreases,
indicating higher confidence in these embeddings.

\begin{figure}[htbp]
\centering
\resizebox{1\linewidth}{!}{%
\begin{tikzpicture}[
    node distance=1.1cm,
    every node/.style={
        align=center, draw, rectangle, rounded corners,
        minimum width=6cm, minimum height=0.8cm
    }
]
    \node[fill=red!20] (hard) {Hard Sample (Large Loss)};
    \node[fill=red!10, below of=hard] (grad_h) {Gradient Backpropagation};
    \node[fill=red!5, below of=grad_h] (scale_h) {Decrease $s_u$};
    \node[fill=red!5, below of=scale_h] (soft_h) {Soften Softmax Distribution};
    \node[fill=red!5, below of=soft_h] (var_h) {Increase $\mathbf{\Sigma}^\text{s}$};

    \node[fill=blue!20, right=2cm of hard] (easy) {Easy Sample (Small Loss)};
    \node[fill=blue!10, below of=easy] (grad_e) {Gradient Backpropagation};
    \node[fill=blue!5, below of=grad_e] (scale_e) {Increase $s_u$};
    \node[fill=blue!5, below of=scale_e] (soft_e) {Sharpen Softmax Distribution};
    \node[fill=blue!5, below of=soft_e] (var_e) {Decrease $\mathbf{\Sigma}^\text{s}$};

    \draw[->, thick, red] (hard) -- (grad_h);
    \draw[->, thick, red] (grad_h) -- (scale_h);
    \draw[->, thick, red] (scale_h) -- (soft_h);
    \draw[->, thick, red] (soft_h) -- (var_h);

    \draw[->, thick, blue] (easy) -- (grad_e);
    \draw[->, thick, blue] (grad_e) -- (scale_e);
    \draw[->, thick, blue] (scale_e) -- (soft_e);
    \draw[->, thick, blue] (soft_e) -- (var_e);

\end{tikzpicture}%
}
 
\caption{Illustration of gradient-driven scale adaptation.}
\vspace{-3mm}
\label{fig:loss_scale}
\end{figure}
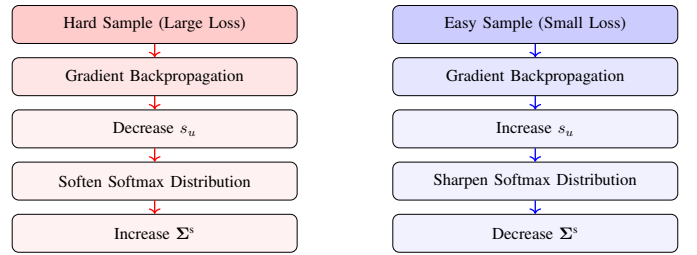

In summary, \textbf{hard samples are associated with smaller scales and larger
variances, while easy samples yield larger scales and lower variances}.
This scale–variance coupling allows uncertainty to be learned through
gradient-driven loss minimization, as shown in Fig. \ref{fig:loss_scale}. 

\section{Uncertainty Estimation Module}
\begin{figure}[tbp]
\vspace{-3mm}
    \centering
    \includegraphics[width=0.9\linewidth]{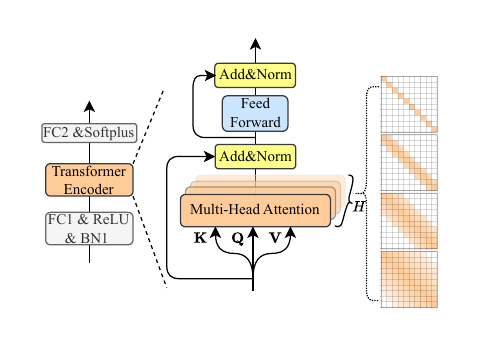}
    \vspace{-8mm}
    \caption{The proposed uncertainty estimation module. Temporal relationships are modeled by a Transformer encoder, and each attention head is assigned with a distinct temporal window, offering multi-view perspectives for estimating frame-level uncertainty.  }
    \vspace{-4mm}
    \label{fig:multihead}
\end{figure}

The previous two sections focus on modifying the training strategies without altering the model architecture. In this section, we aim to redesign the uncertainty estimation module. In our initial work \cite{lee2021xi}, the uncertainty estimation relies on two simple learnable layers with basic activation functions to predict frame-level uncertainty, as illustrated in Fig. \ref{fig:xi}. However, this design is relatively  shallow and lacks the ability to capture temporal dependencies across speech frames. In our previous work \cite{li2025xi+} addressed this limitation by incorporating a Transformer encoder to  model temporal relations, thereby enhancing the model’s capacity to capture complex non-linear patterns. 
Although Transformer-based architectures have been explored for speaker recognition, they often struggle to adequately capture short-range local context \cite{sang2023improving,han2022local,wang2022multi,tu2025confusionformer}. Among the proposed variants, multi-view self-attention (MVA) \cite{wang2022multi} is particularly noteworthy: each attention head is assigned a distinct temporal window, enabling it to focus on different receptive-field ranges.  Such a property aligns well with our goal, as it allows the uncertainty estimation module to effectively capture both long-range dependencies and fine-grained local patterns, which are essential for  robust and reliable uncertainty estimation, as shown in Fig. \ref{fig:multihead}. For each attention head, the receptive-field size is defined as:
\begin{equation}
w_h = 2^{(h+1)} + 1,
\end{equation}
where $h \in \{0, \ldots, H-1\}$ denotes the head index among a total of $H$ heads, and $w_h$ represents the temporal window size for the $h$-th head.

 \vspace{-3mm}
\section{Experimental setup}

\subsection{Dataset}
All models are trained on VoxCeleb2 \cite{chung2018voxceleb2} and evaluated on the in-domain VoxCeleb1 test set \cite{nagrani2017voxceleb}, as well as on cross-domain test sets including Speakers in the Wild (SITW) \cite{mclaren2016speakers} and CNCeleb \cite{fan2020cn} \footnote{For SITW, the core–core condition is used. For CNCeleb, the evaluation list from CNC-Eval-Avg in WeSpeaker \cite{wang2023wespeaker} is adopted. }.
Data augmentation is applied during training using additive noise from the MUSAN corpus \cite{snyder2015musan} and simulated reverberation via room impulse responses from the RIR database \cite{ko2017study}.

\subsection{Training Strategy}

We adopt the training pipeline from the VoxCeleb v2 recipe provided by the WeSpeaker toolkit\footnote{\url{https://github.com/wenet-e2e/wespeaker/blob/ master/examples/voxceleb/v2/run.sh}} \cite{wang2023wespeaker}, following all default hyperparameter settings. The training spans 150 epochs, using 2-second audio segments. The default scale $s$ in AAM-softmax and  is 32. The angular margin is initially set to zero and then progressively increased from 0 to 0.2 between epochs 20 and 40, after which it remains fixed. Finally, we average the parameters from the last 10 checkpoints to obtain a final model checkpoint.  Unless otherwise specified, ECAPA-TDNN \cite{desplanques20_interspeech} is adopted as the default speaker encoder.

\subsection{Evaluation Protocol }

We report the performances in terms of the equal error rate (EER) and the minimum detection cost function (minDCF) with P$_\text{target}$ = 0.01 and C$_\text{FA}$ = C$_\text{Miss}$ = 1. The scores are produced by calculating the cosine distance between embeddings.

\section{Results and Discussions}
\renewcommand{\arraystretch}{0.95}
\begin{table*}[bp]
\vspace{-3mm}
    \centering
    \caption{Overall results on Voxceleb1 in terms of the EER (\%) and minDCF (\textit{P}target = 0.01) where lower values indicate better performance.   All experiments in this table employ ECAPA-TDNN \cite{desplanques20_interspeech} as the speaker encoder. $\Delta$ denotes relative improvement compared to the benchmark.}
    \begin{tabular}{c|c|c|c|c|c|c|c|c|c|c|c|c}
    \hline
        \multirow{2}{*}{\#Exp.} & \multirow{2}{*}{Model} & \multirow{2}{*}{\#Param.}   & \multicolumn{2}{c|}{$\mathcal{L}_\text{SVL}$} & Uncertainty & \multicolumn{2}{c|}{Vox1-O} & \multicolumn{2}{c|}{Vox1-E} & \multicolumn{2}{c|}{Vox1-H} & Avg.  \\ \cline{7-12} \cline{4-5} 
        & & & Method & $\lambda$ &Scoring $\rho$ & EER  & minDCF & EER & minDCF & EER & minDCF & $\Delta$\\ \hline 
        1  \footnotemark & x-vector \cite{desplanques20_interspeech}& 6.19 M& \multirow{3}{*}{\ding{56} } & \multirow{3}{*}{0}&      \multirow{2}{*}{0 } & 1.069 & 0.122 & 1.209 & 0.136 & 2.310 & 0.226 & -\\ \cline{1-3} \cline{7-12} 
        \multirow{2}{*}{2} & \multirow{2}{*}{xi-vector \cite{lee2021xi}} & \multirow{2}{*}{5.90 M} &  &   &  &  0.995 & 0.103 & 1.130 & 0.126 & 2.169 &0.209 & Benchmark\\ \cline{6-13}
        & & & & &  $1/d$ \cite{wang2024cosine}& 0.989 &0.100 &1.123 &0.125 &2.158 &0.208 & 1.00\%  \\ \hline
       
        \multirow{3}{*}{3} & \multirow{9}{*}{xi-vector+$\mathcal{L}_\text{SVL}$} & \multirow{9}{*}{5.90 M} &  \multirow{9}{*}{$\mathcal{L}_\text{SVL-Fix}$} & \multirow{3}{*}{0.01} & 0 & 1.048 &0.126 &1.142 &0.128 &2.202 &0.205 & -4.95\%\\
        & & & & & $1/d$ &  1.053 & 0.124 & 1.137 &0.128 & 2.195 & 0.204 & -4.57\% \\
        & & & & & $\alpha (0.0863)$ & 0.994 &0.103 &1.080 &0.117 &2.088 &0.201& 3.21\% \\ \cline{1-1} \cline{5-13}

        \multirow{3}{*}{4} &  &  &   & \multirow{3}{*}{0.05} & 0 & 1.021 &0.108 &1.139 &0.128 &2.176 &0.205 & -1.71\%\\
        & & & & & $1/d$ & 1.005 & 0.107 & 1.135 & 0.127 & 2.164 & 0.204 &-0.58 \\
           & & & & & $\alpha(0.0848)$ & 0.936 & 0.100 &\textbf{1.076} & 0.116 &2.068 &0.202 & 4.92\%\\ \cline{1-1} \cline{5-13}

        \multirow{3}{*}{5} &  &  &  & \multirow{3}{*}{0.1} & 0 & 1.042 &0.115 &1.150 &0.127 &2.204 & 0.209&-3.43\% \\
        & & & & & $1/d$ &1.042 & 0.114 & 1.148 & 0.127 & 2.200 & 0.205&-2.88\%   \\
        & & & & & $\alpha(0.0995)$ & 0.994 &0.101 &1.103 &0.119 &2.113 &\textbf{0.198}&2.98\%\\ \hline

        \multirow{3}{*}{6} & \multirow{9}{*}{xi-vector+$\mathcal{L}_\text{SVL}$} & \multirow{9}{*}{5.90 M} &  \multirow{9}{*}{$\mathcal{L}_\text{SVL-Pro}$} &\multirow{3}{*}{0.01} & 0 & 1.064 & 0.114& 1.157 & 0.127 & 2.163 & 0.211 & -3.57\% \\
        & & & & & $1/d$ &1.053 & 0.111 & 1.150 & 0.126 & 2.154 & 0.210 &-2.53\%    \\
        & & & & & $\alpha(0.0749)$ & \textbf{0.931} & 0.102 & 1.079 & \textbf{0.115} & \textbf{2.046}& 0.201 & 5.02\%  \\ \cline{1-1} \cline{5-13}

        \multirow{3}{*}{7} &  & &   &\multirow{3}{*}{0.05} & 0 & 1.064 &0.117 &1.160& 0.125 &2.155 &0.210 & -3.70\%\\
        & & & & & $1/d$ &1.063 & 0.115 & 1.155 & 0.125 & 2.148& 0.208  & -2.74\%  \\
        & & & & & $\alpha(0.0766)$ & 0.952 &\textbf{0.099} &1.095 &0.117 &2.053 &\textbf{0.198}& 4.84\% \\ \cline{1-1} \cline{5-13}

        \multirow{3}{*}{8} &  & &   &\multirow{3}{*}{0.1} & 0 & 1.032& 0.117 &1.161 &0.126 &2.161 &0.212 &-3.52\% \\
        & & & & & $1/d$ & 1.031 & 0.115 & 1.158 & 0.125 & 2.153 & 0.211  &-3.22\%\\
        & & & & & $\alpha(0.0759)$ &0.952 &0.100 &1.110 &0.117 &2.065 & 0.200 &4.22\% \\ \hline

    \end{tabular}
    \label{tab:SVL}
\end{table*}
\footnotetext{Results come from \url{https://github.com/wenet-e2e/wespeaker/blob/master/examples/voxceleb/v2/README.md}}

\subsection{Visualization of Uncertainty Estimation}

Unlike previous temporal pooling layers that rely on statistical operators or attention mechanisms, we explicitly estimate the covariance of each frame based on Gaussian posterior inference and use it as an importance weight. A larger covariance indicates that the corresponding frame is less reliable and contributes less to the final utterance-level speaker embedding.

\begin{figure}[tbp]
    \centering
\vspace{-3mm}
    \subfloat[Frame-wise precision under varying SNR conditions. 
    The SNR levels for the intervals 0–10 s, 10–20 s, 20–30 s, 
    30–40 s, 40–50 s, and 50–60 s are clean, 20 dB, 10 dB, 
    0 dB, –10 dB, and –20 dB, respectively.]{
        \includegraphics[width=0.9\linewidth]{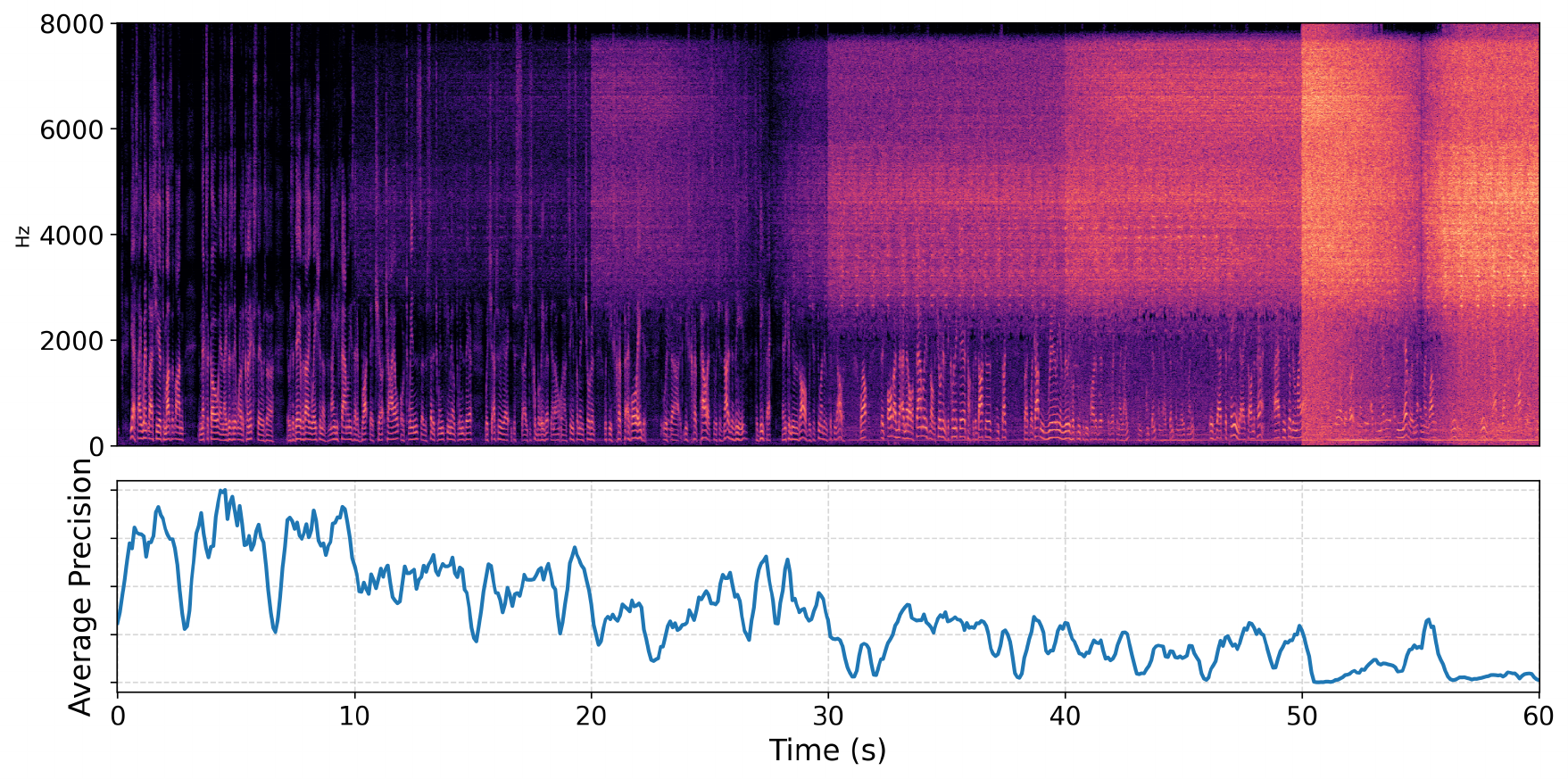}
        \label{fig:precision_snr}
    }
    \vspace{-3mm}

    \subfloat[Frame-wise precision under different spectrogram 
    patch masking levels. From left to right, increasingly 
    larger portions of the spectrogram are masked.]{
        \includegraphics[width=0.9\linewidth]{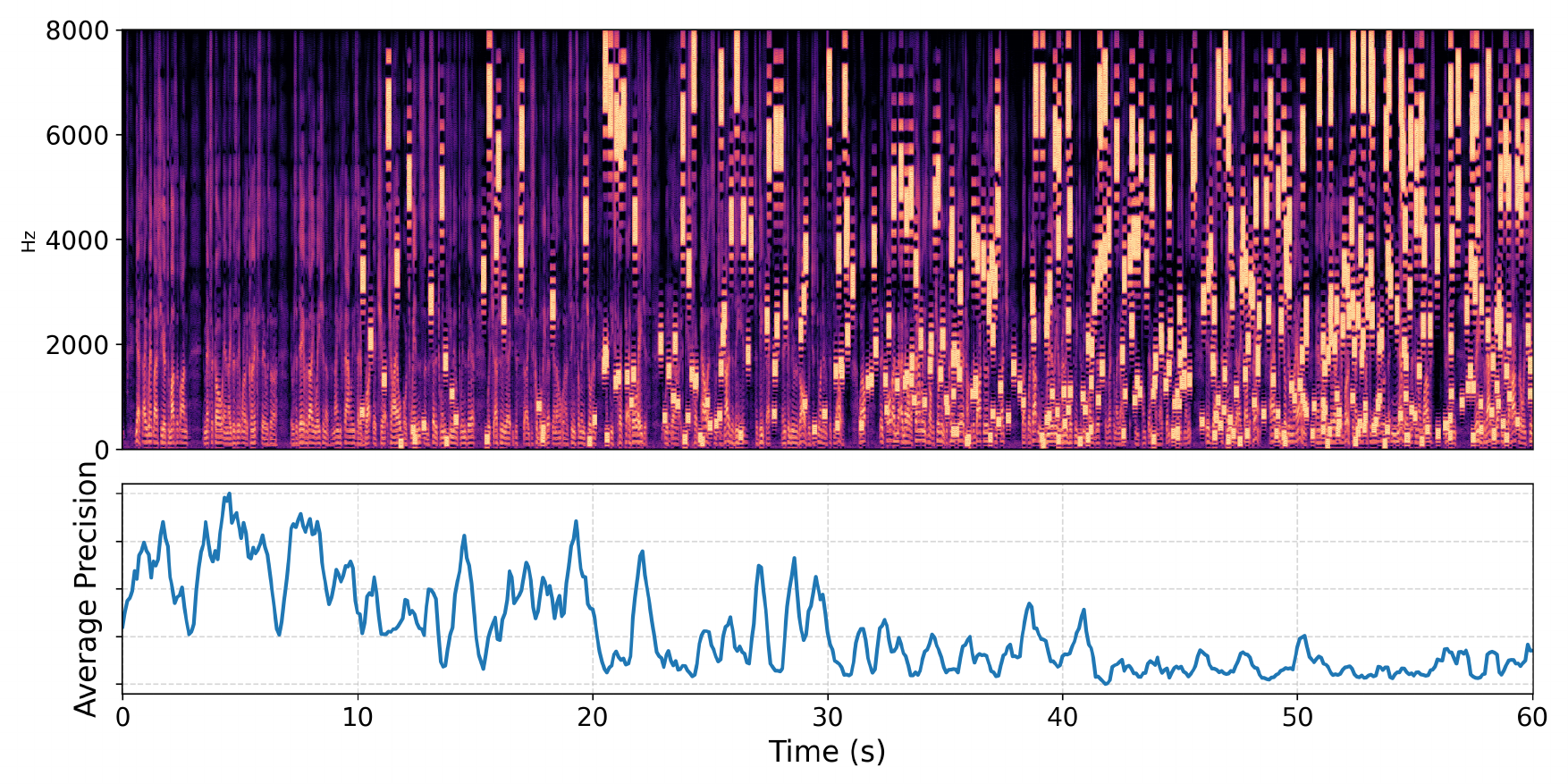}
        \label{fig:precision_patch}
    }

    \caption{Comparison of frame-wise precision under (a) noise corruption and 
    (b) spectrogram masking conditions.}
    \label{fig:precision_visualization}
    \vspace{-3mm}
\end{figure}

To verify that the estimated covariance indeed aligns with our objective, we visualize the temporal variation of the precision matrix (i.e., the inverse of the covariance matrix). Specifically, we average the precision matrix along the channel dimension and plot its values over time, as shown in Fig.~\ref{fig:precision_visualization}. The input utterance is 60 seconds long, and we apply two types of distortions: adding noise and masking partial regions of the spectrogram.
From the two visualizations, we observe that the precision values decrease as the speech becomes  severely corrupted. Moreover, even regions containing weak or sparse speech exhibit lower precision, indicating reduced reliability. These observations confirm that the estimated frame-level uncertainty effectively reflects the quality and trustworthiness of the speech signal.

\subsection{The Effectiveness of Speaker-level Uncertainty Supervision}

We have shown that the frame-level uncertainty estimation within an utterance is
accurate. However, the model applies a $softmax$ function to normalize uncertainty
scores within each utterance, which preserves only the internal ordering of
frame-wise uncertainty while eliminating the absolute scale. As a result,
uncertainty values across different utterances become incomparable, and
$\mathbf{\Sigma}^\text{s}$ cannot be reliably interpreted or supervised at the
utterance level. To address this issue, we propose a speaker-level uncertainty
supervision mechanism and verify its effectiveness in this section.

\begin{table*}[hbp]
\vspace{-3mm}
    \centering
    \caption{Overall results of global-level uncertainty supervision on in-domain and cross-domain sets. $\Delta_{\cos} = (\cos\theta_{y_i} - \displaystyle\max_{j \neq y_i}  \cos\theta_j).\text{detach()}$ which serves as the difficulty indicator.}
     
    \begin{tabular}{cccc|cccc|ccc}
    \toprule 
       \multirow{4}{*}{\# Exp}   & \multirow{4}{*}{Loss} & \multirow{4}{*}{$\mathbf{\Lambda}$} & \multirow{4}{*}{$\mathbf{\Sigma^s}$} & \multicolumn{4}{c|}{In-domain Test} &  \multicolumn{3}{c}{Cross-domain Test} \\ 
       & & & & Vox1-O & Vox1-E & Vox1-H & Avg. &SITW-Dev& SITW-Eval & Avg.    \\ \cline{5-7} \cline{9-10}
       & & & & EER & EER & EER & \multirow{2}{*}{$\Delta$}&  EER & EER& \multirow{2}{*}{ $\Delta$} \\
       & &  &  & minDCF & minDCF & minDCF & &  minDCF & minDCF  \\ \hline \hline 
       \multirow{2}{*}{2}  &  \multirow{2}{*}{$\mathcal{L}_\text{AAM}$} &  \multirow{2}{*}{-} &  \multirow{2}{*}{\ding{52}} &  \textbf{0.995} & 1.130 & 2.169 &\multirow{2}{*}{Benchmark} &  2.079 & 2.357 & \multirow{2}{*}{Benchmark}\\ 
       & & & &  \textbf{0.103} & 0.126 & 0.209 &  & 0.180 & \textbf{0.175} &\\\hline 
        \multirow{2}{*}{9}  &  \multirow{2}{*}{$\mathcal{L}_{\text{UAAM}}$} &  \multirow{2}{*}{$\mathbf{I}$} & \multirow{2}{*}{\ding{52}} & 1.090&1.159 &2.188 & -4.33\% & 2.079 & 2.314& 0.91\% \\ 
        & &&&  0.120 &0.125&0.209&-5.28\% &0.171 & 0.182&0.79\%  \\ \hline 
            \multirow{2}{*}{10}  &  \multirow{2}{*}{$\mathcal{L}_{\text{UAAM}}$} &   \multirow{2}{*}{$\mathbf{I} - \Delta_{\cos} $}& \multirow{2}{*}{\ding{52}}& 1.010 & 1.120 & 2.107 &0.74\% &2.199& \textbf{1.998}&4.73\%  \\ 
    &&&&  0.108 & 0.126 & 0.205& -1.65\% & 0.171 & 0.181   & 0.79\%\\ \hline
    \multirow{2}{*}{11}  &  \multirow{2}{*}{$\mathcal{L}_{\text{UAAM}}$} &  \multirow{2}{*}{$\mathbf{I} - \Delta_{\cos} $}& \multirow{2}{*}{\ding{56}} & 1.128 & 1.137 & 2.122 & -3.94\% &  2.128& 2.433 &-2.79\% \\
    & & & & 0.106 & 0.124 & 0.210 &-0.60\% &  0.178 & 0.183 & -1.73\% \\ \hline
        \multirow{2}{*}{12}  &  \multirow{2}{*}{$\mathcal{L}_{\text{UAAM}}$} &  \multirow{2}{*}{$0.5\,\mathbf{I} - \Delta_{\cos} $}& \multirow{2}{*}{\ding{52}} & \textbf{0.995} & \textbf{1.087} & \textbf{1.957} &4.53\% &\textbf{2.050} &2.214 & 3.73\%\\ 
        &&&&  0.107 & \textbf{0.120} & \textbf{0.196}& 2.37\% & \textbf{0.158} & 0.177& 5.54\% \\ \hline
        \bottomrule
    \end{tabular}
    \vspace{-3mm}
    \label{tab:global-level}
\end{table*}

\begin{figure}[htbp]
\vspace{-5mm}
    \centering
    \includegraphics[width=1\linewidth]{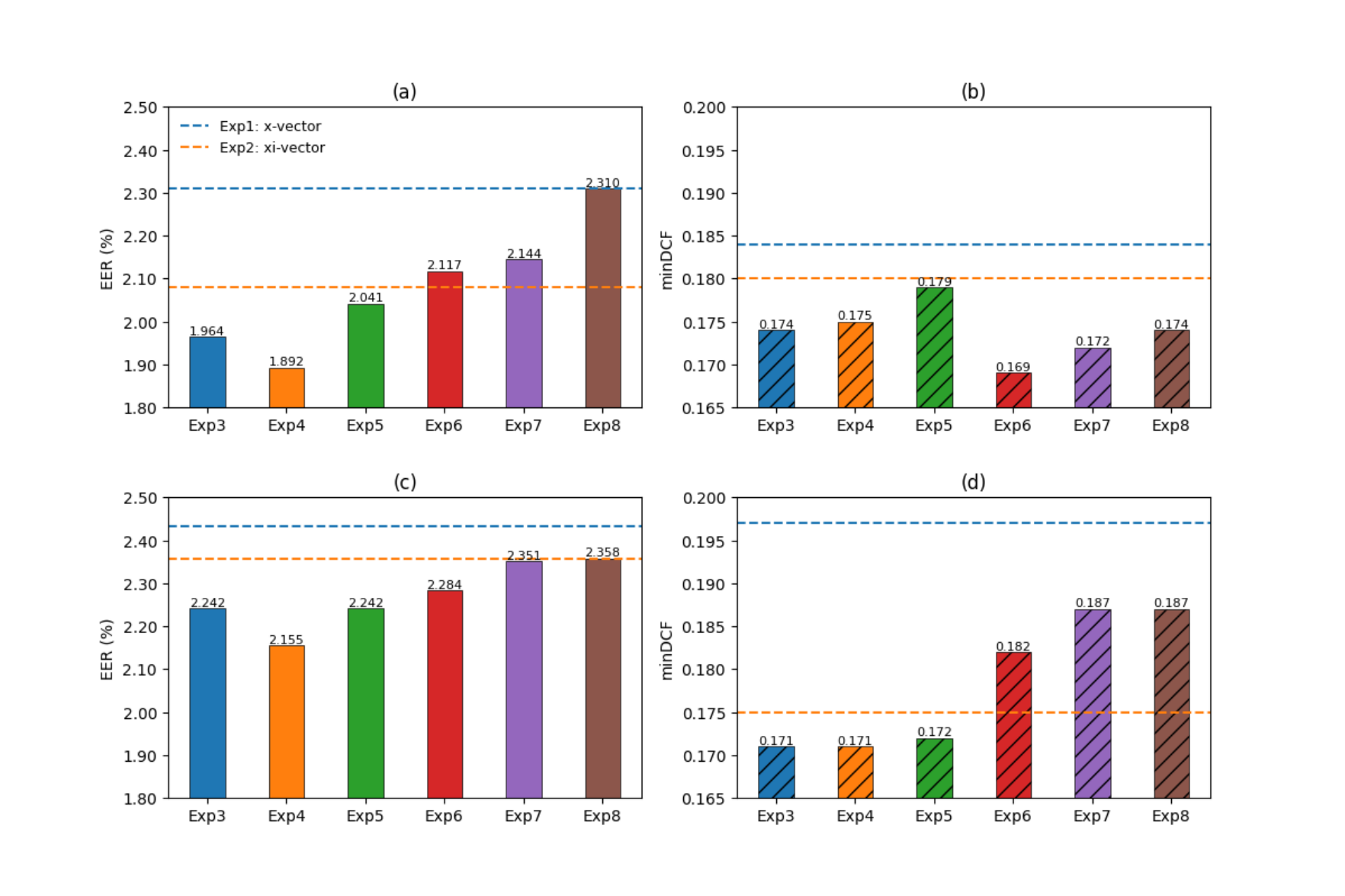}
    \vspace{-8mm}
    \caption{Results of $\mathcal{L}_\text{SVL}$ on cross-domain test set: (a) EER and (b) minDCF on the SITW-Dev set; (c) EER and (d) minDCF on the SITW-Eval set.}
    \label{fig:SVL_CrossTest}
    \vspace{-5mm}
\end{figure}

As shown in Table~\ref{tab:SVL}, we compare the performance of our two \textit{Stochastic Variance Loss} variants, $\mathcal{L}_\text{SVL-Fix}$ and $\mathcal{L}_\text{SVL-Pro}$. The results indicate that using $\mathcal{L}_\text{SVL}$ alone does not yield noticeable improvements. However, combining it with the \textit{uncertainty-aware} cosine scoring strategy leads to better performance. We further evaluate two strategies for the scaling factor $\rho$: (1) the static factor $1/d$ proposed by Wang~\cite{wang2024cosine}, and (2) a learnable factor $\alpha$ trained jointly with the model proposed by our previous work xi+ \cite{li2025xi+}. Results demonstrate that the learnable scaling factor significantly outperforms the static counterpart. 
When combining $\mathcal{L}_{\text{SVL}}$ with the learnable scaling factor $\alpha$, consistent and significant improvements are observed across different loss weights $\lambda$. We further evaluate the models on a cross-domain test set, as shown in Fig.~\ref{fig:SVL_CrossTest}. Exps. 3–5 consistently outperform the two baseline methods across all metrics, whereas Exps. 6–8 exhibit unstable behavior and, in some cases, even underperform compared to the baselines. These results indicate that $\mathcal{L}_{\text{SVL-Fix}}$ is more effective, providing stronger robustness and better generalization to cross-domains. Additionally, Exp. 4 with $\lambda=0.05$ achieves the best overall performance.

\subsection{The Effectiveness of Global-level Uncertainty Supervision}
Speaker-level supervision constrains uncertainty only within utterances from the same speaker and cannot compare uncertainty across different speakers. In practice, however, each training sample produces a different loss value, reflecting its varying difficulty. These loss differences are back-propagated to the uncertainty-aware scale $s_u$ in $\mathcal{L}_{\text{UAAM}}$, causing the scale to update dynamically.  In this section, we explore the effectiveness of the proposed global-level uncertainty supervision.

\begin{table*}[htp]
    \centering
    \vspace{-3mm}
    \caption{Comparison of different attention variants for uncertainty estimation. `–4' denotes the use of a 4-head attention configuration. The speaker encoder across all experiments is still Ecapa-tdnn. }
     
    \begin{tabular}{lccccc|cc|c}
    \hline
    \toprule 
       \multirow{3}{*}{Model}  & \multirow{2}{*}{\# Param}. & \multirow{2}{*}{Flops} & \multicolumn{3}{c|}{In-domain Test} & \multicolumn{2}{c|}{Cross-domain Test} \\ \cline{4-8}
       & & & Vox1-O & Vox1-E & Vox1-H  &SITW-Dev& SITW-Eval & Avg.    \\ \cline{4-8}
       & (M) & (G)& EER / minDCF & EER / minDCF & EER / minDCF & EER / minDCF & EER / minDCF & $\Delta$ \\ \hline \hline 
        xi-vector \cite{lee2021xi}&  5.90  & 1.04 & 0.995 / \textbf{0.103}& 1.130 / \textbf{0.126}& 2.169 / 0.209& 2.079 / 0.180 & 2.357 / 0.175 &Benchmark  \\ \hline
        xi+MHA-4  & 6.69  & 1.20 & 1.016 / 0.122 & 1.170 / 0.133 & 2.223 / 0.219 & 2.079 / 0.177 & 2.264 / 0.182 &-3.53\% \\ 
        xi+MHA-8  & 6.69 & 1.20 & 1.048 / 0.116 & 1.145 / 0.128 & \textbf{2.117} / 0.207 & 1.733 / 0.165 & 1.941 / 0.181 &2.17\%  \\ 

    xi+MHA-16  & 6.69  & 1.20 &  \textbf{0.872} / 0.126 & 1.128 / 0.127 & 2.127 / 0.211 &1.771 / 0.168 &2.012 / 0.202 &1.11\%\\ 
    xi+MHA-32  & 6.69  & 1.20 &  1.032 / 0.145 & 1.127 / \textbf{0.126} & 2.122 / 0.217 & 1.810 / 0.164 & 1.919 / 0.171 & -0.32\% \\ \hline 
        
        xi+MVA-8 & 6.69  & 1.20 &  1.010 / 0.107 & \textbf{1.123} / 0.128 & 2.121 / \textbf{0.206} & \textbf{1.617} / \textbf{0.159} & \textbf{1.867} / \textbf{0.170} & 5.48\%\\ \hline
        \bottomrule
    \end{tabular}
    
    \label{tab:model}
    \vspace{-3mm}
\end{table*}

\begin{figure}[htbp]
    \centering
    \includegraphics[width=0.9\linewidth]{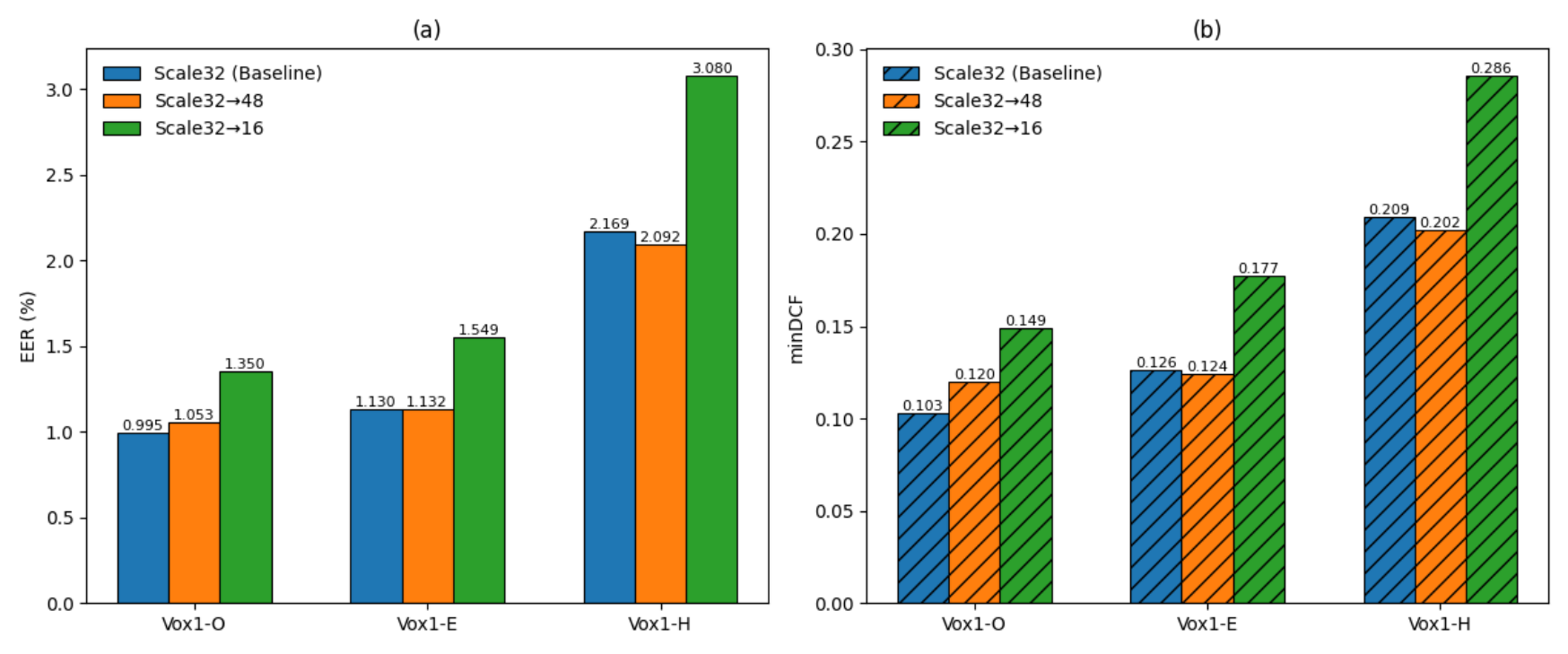}
    \vspace{-3mm}
    \caption{Speaker verification performance under different scale settings. (a) EER across different scale settings. (b) minDCF across different scale settings. }
    \label{fig:scale}
    \vspace{-3mm}
\end{figure}
\subsubsection{Preliminary Analysis of Scale Effects}
\label{sec:scale}
As far as we know, few studies have systematically examined the effect of the scale parameter in speaker recognition.
In this section, we investigate the effect of the scale $s$ and how varying its value influences the performance of the speaker recognition system. The default scale in AAM-Softmax is $s=32$. Based on this setting, we examine how the performance changes when the scale is increased or decreased.
Specifically, we design two additional experiments:
(1) a linearly increasing schedule where the scale grows from 32 to 48, and
(2) a linearly decreasing schedule where the scale drops from 32 to 16.
As shown in Fig.~\ref{fig:scale}, increasing the scale generally improves performance on the hard test set (Vox1-H), while yielding slightly worse results on the easier sets (Vox1-O and Vox1-E). This behavior arises because a larger scale sharpens the softmax distribution, causing the gradient to emphasize harder samples.
 In contrast, decreasing the scale consistently degrades performance across all test sets, indicating that decreasing  scale severely harms discriminability.

\subsubsection{Ablation Study on the Bias Term $\mathbf{\Lambda}$}

Table \ref{tab:global-level} reports the performance of our proposed uncertainty-aware AAM-Softmax ($\mathcal{L}_{\text{UAAM}}$ \ref{eq:UAAM}). Exp. 9 adopts a static constant Identity matrix as the bias term $\mathbf{\Lambda}$, since the covariance matrix $\mathbf{\Sigma}^\text{s} \geq0$, the resulting uncertainty-aware scale $s_u$ is always less than 1. Consequently, the final scale applied during training  $s\cdot s_u$ is less than the original scale $s$.  According to our analysis in the previous section, it is detrimental to performance. As expected, Exp. 9 fails to outperform the baseline.

Exp.~10 adopts a dynamically varying bias term $\mathbf{\Lambda}$ derived from
$
\Delta_{\text{cos}} = \big(\cos\theta_{y_i} - \max_{j \neq y_i} \cos\theta_j\big).\texttt{detach()}$,
which measures the cosine margin between the target class and the most competitive non-target class.
The \texttt{detach()} operation prevents gradients from flowing through $\Delta_{\text{cos}}$,
ensuring that it serves solely as a difficulty indicator rather than a trainable quantity.
Intuitively, $\Delta_{\text{cos}}$ reflects the sample-level classification difficulty:
larger values indicate that the target class is already well separated from non-target classes,
corresponding to easier samples, whereas smaller or negative values imply harder samples.
Empirically, we observe that most $\Delta_{\text{cos}}$ values fall within the range of $(-0.8,\,0.5)$.

We assume that $\Delta_{\text{cos}}$ plays two important roles:

\begin{enumerate}[label=\arabic*.]
    \item   $\Delta_{\text{cos}}$ tends to increase as training proceeds, which naturally drives the uncertainty scale $s_u$ upward in later stages. This behavior aligns with our intuition in Section \ref{sec:scale}, that a larger scale is desirable. 
    \item  At the early stage of training, the estimated uncertainty $\mathbf{\Sigma}^\text{s}$ is neither accurate nor sufficiently discriminative across different samples. Incorporating $\Delta_{\text{cos}}$ introduces sample-specific variation into $s_u$ for different samples, providing a learning signal that helps the model form more meaningful uncertainty estimates.
\end{enumerate}

Exp. 10 achieves slightly better performance than Exp. 9, demonstrating the benefit of integrating $\Delta_{\text{cos}}$. Exp. 12 further modifies the design by setting $\mathbf{\Lambda} = 0.5\,\mathbf{I} - \Delta_{\text{cos}}$\footnote{$0.5\,\mathbf{I}$ is the smallest constant we tested that prevents NaN issues during training while ensuring that $\mathbf{\Lambda}$ remains non-negative.}. This reduces the constant component and increases the dynamic influence of $\Delta_{\text{cos}}$. As a result, Exp. 12 outperforms Exp. 10 in terms of average performance, particularly on the in-domain test sets.

To verify the necessity of modeling utterance-level uncertainty, we ablate the covariance term by removing $\mathbf{\Sigma}^\text{s}$ and retaining only $\mathbf{\Lambda}$ in (\ref{equ:uncertainty}), as shown in Exp. 11. Compared with Exp. 10, this removal leads to a clear performance drop, indicating that estimating utterance-level uncertainty is essential.
This result indicates that the gains from global-level uncertainty supervision are not solely due to the dynamic scaling effect, but also arise from more accurate, sample-specific uncertainty estimation provided by $\mathbf{\Sigma}^\text{s}$.

\subsection{The Effectiveness of Temporal Modeling}
In this section, we incorporate a Transformer encoder to model temporal dependencies and compare multi-head attention (MHA) with different head configurations, as reported in Table~\ref{tab:model}. The results show that the 8-head configuration achieves the best performance among all MHA variants. Based on this setting, we further encourage each attention head to capture a distinct temporal context, forming the multi-view self-attention (MVA) mechanism. Experimental results demonstrate that MVA further improves performance, yielding over a 5\% relative gain while introducing only minimal additional parameters and computational overhead.

\begin{table*}[htp]
\vspace{-3mm}
    \centering
    \caption{Ablation study of progressively incorporating  each component in the proposed $\mathcal{U}^3$-xi model.}
    \begin{tabular}{clcccc|cc|c}
    \hline
    \toprule 
       \multicolumn{2}{c}{\multirow{3}{*}{Model}} &\multirow{3}{*}{\#Param.}  & \multicolumn{3}{c|}{In-domain Test} & \multicolumn{2}{c|}{Cross-domain Test} \\ \cline{4-8}
       &&& Vox1-O & Vox1-E & Vox1-H  &SITW-Dev& SITW-Eval & Avg. $\Delta$   \\ \cline{4-8}
      & & & EER / minDCF & EER / minDCF & EER / minDCF & EER / minDCF & EER / minDCF &  EER / minDCF\\ \hline \hline 
       \multicolumn{2}{c}{ASTP \cite{desplanques20_interspeech}}  &6.19 M &  1.069 / 0.122 & 1.209 / 0.136 & 2.310 / 0.226 & 2.310 / 0.184 & 2.433 / 0.197 & Benchmark \\ \hline
        \multicolumn{2}{c}{xi \cite{lee2021xi}}&5.90 M& 0.995 / 0.103 & 1.130 / 0.126 & 2.169 / 0.209 & 2.079 / 0.180 & 2.357 / 0.175 & 6.53\% / 8.76\% \\ \hline
      \multirow{3}{*}{$\mathcal{U}^3$-xi}& +MVA-8 &6.69 M &  1.010 / 0.107 & 1.123 / 0.128 & 2.121 / 0.206 & \textbf{1.617} / \textbf{0.159} & \textbf{1.867} / 0.170 & 14.81\% / 10.86\% \\ 
       & \quad+$\mathcal{L}_{\text{UAAM}}$ & 6.69 M &   0.856 / 0.109 & 1.064 / 0.121  & 1.982 / 0.195  & 1.854 / 0.162 & 2.105 / 0.170 & 15.85\% / 12.22\% \\  
        &  \quad \quad + $\rho=1$ & 6.69 M & \textbf{0.782} / \textbf{0.100} & \textbf{1.016} / \textbf{0.115} & \textbf{1.888} / \textbf{0.187} & 1.770 / 0.160 & 1.921 / \textbf{0.169} &21.10\% / 15.57\% \\ \hline
        \bottomrule
    \end{tabular}
    \label{tab:U_cube}
    \vspace{-3mm}
\end{table*}
 
\subsection{$\mathcal{U}^3$-xi: Combining All Components}

In this section, we integrate all proposed components into a unified model, including the global-level uncertainty supervision ($\mathcal{L}_{\text{UAAM}}$), the MVA uncertainty estimation module, and the uncertainty-aware scoring mechanism. This combined system, referred to as $\mathcal{U}^3$-xi, allows us to examine the cumulative benefit and the interaction among these three complementary designs. Table \ref{tab:U_cube} presents the performance improvements obtained by progressively incorporating each component. To more clearly verify the effectiveness of the proposed uncertainty estimation, we use the original ECAPA-TDNN with attentive statistical temporal pooling (ASTP) \cite{desplanques20_interspeech} as the baseline.

We progressively add the proposed modules as follows:
\begin{enumerate}[label=\arabic*.]
    \item Incorporating uncertainty:
The original xi-vector \cite{lee2021xi} replaces ASTP with a two-layer Gaussian posterior inference module that estimates frame-level uncertainty using two simple linear projections. This architectural change alone yields relative improvements of 6.53\% in EER and 8.76\% in minDCF, demonstrating the effectiveness of  incorporating uncertainty estimation into the speaker embedding framework.
\item  Introducing MVA for temporal modeling:
Building on the previous configuration, we add a MVA module  to enhance nonlinear modeling capacity and capture richer temporal dependencies. This addition yields further relative performance gains, demonstrating the effectiveness of strengthened temporal uncertainty modeling. 
\item Applying global-level uncertainty supervision:
By inserting uncertainty into the scale of $softmax$ function, $\mathcal{L}_{\text{UAAM}}$, the model receives explicit supervision to learn more reliable uncertainty representations. This global guidance further enhances generalization across both in-domain and cross-domain conditions. 
\item  Incorporating uncertainty scoring:
We initially attempted to learn the scaling factor $\rho$ using the SVL loss
in Eq.~\ref{equ:new_svl}. However, this scheme failed to converge, with NaN
values  observed, indicating an optimization conflict between
$\mathcal{L}_{\text{SVL}}$ and $\mathcal{L}_{\text{UAAM}}$.
We therefore fix $\rho = 1$ as a static scaling factor. Despite its simplicity,
this setting still yields performance gains.
This behavior is reasonable, as the proposed global-level uncertainty
supervision already enforces discriminative utterance-level uncertainty,
making additional learnable scaling unnecessary.
With uncertainty-aware scoring enabled, the final system achieves relative
improvements of 21.1\% and 15.57\% over the baseline in terms of EER and
minDCF, respectively.

\end{enumerate}

\subsection{Systematic Analysis of Uncertainty in the $\mathcal{U}^3$-xi Model}
 
\begin{figure}[tbp]
    \centering
    \vspace{-3mm}
    \subfloat[xi-vector]{%
        \includegraphics[width=0.49\linewidth]{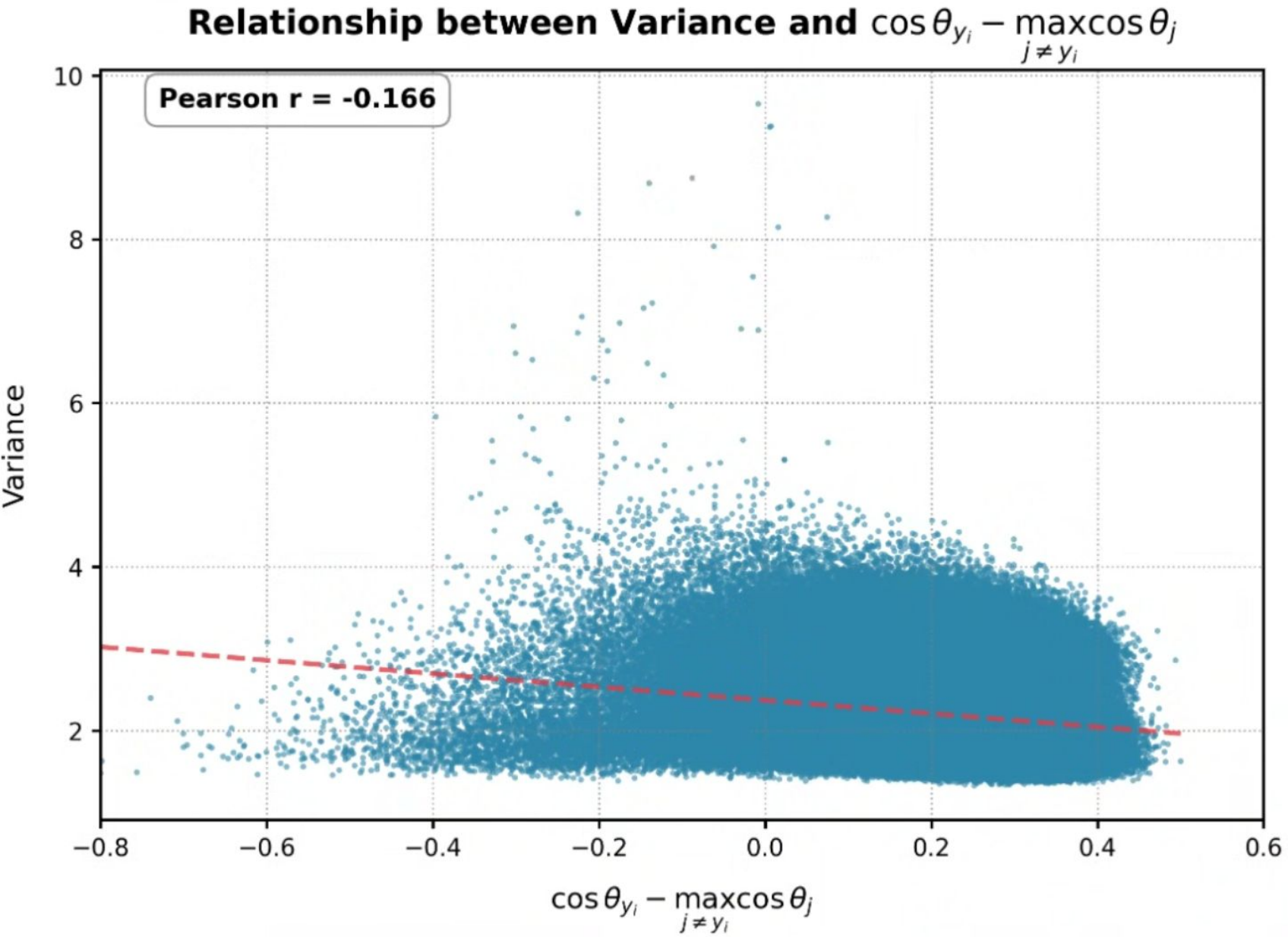}
    }
    \hfill
    \hspace{-2mm}
    \subfloat[$\mathcal{U}^3$-xi]{%
        \includegraphics[width=0.49\linewidth]{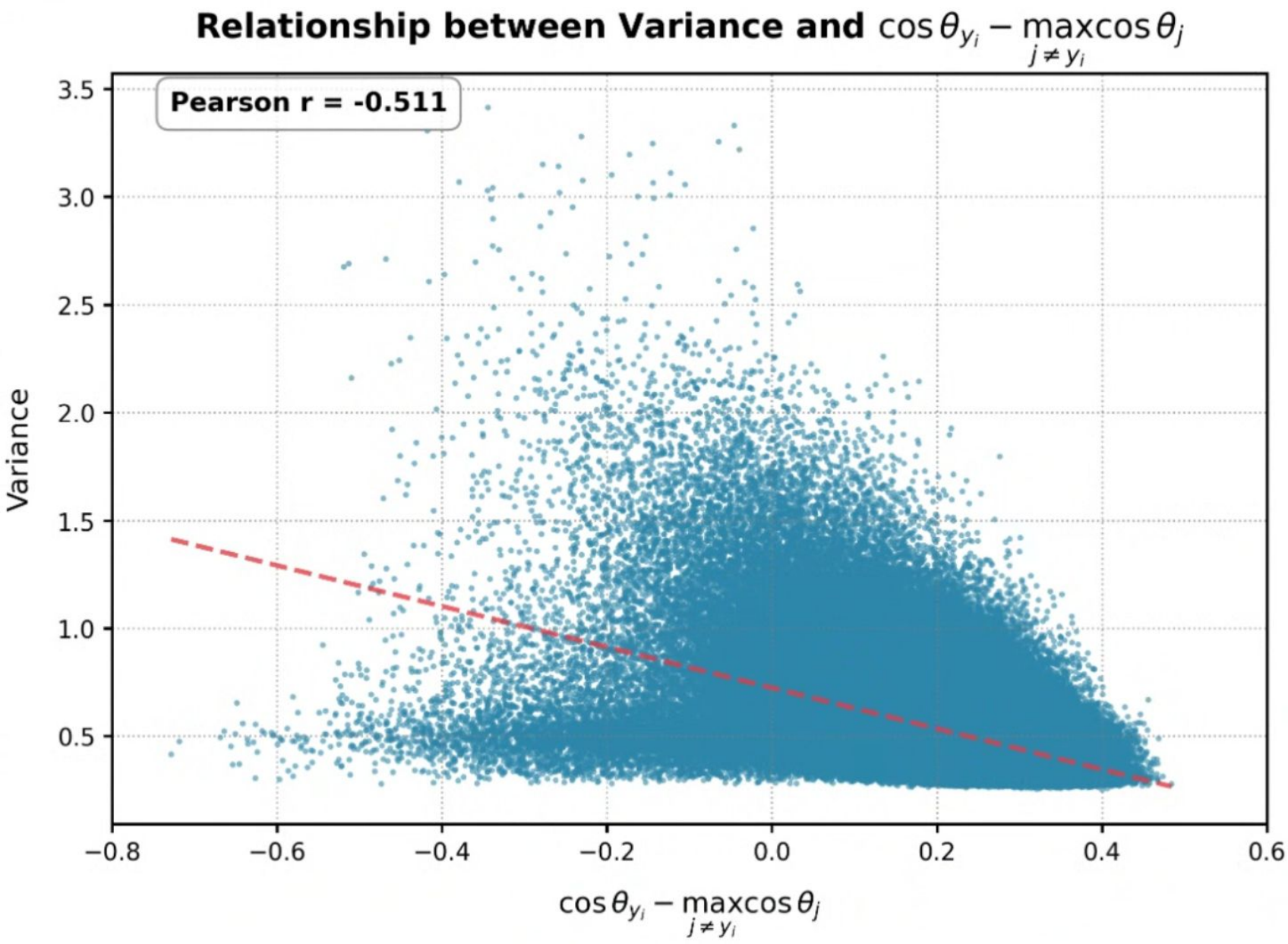}
    }

    \caption{
    Relationship between utterance-level variance and 
    sample difficulty on the training set. 
    The y-axis shows the averaged diagonal variance across all embedding dimensions.
    }\vspace{-3mm}
    \label{fig:variance_softmax}
\end{figure}

\begin{figure}[htbp]
    \centering
    \vspace{-3mm}
    \subfloat[xi-vector]{%
        \includegraphics[width=0.49\linewidth]{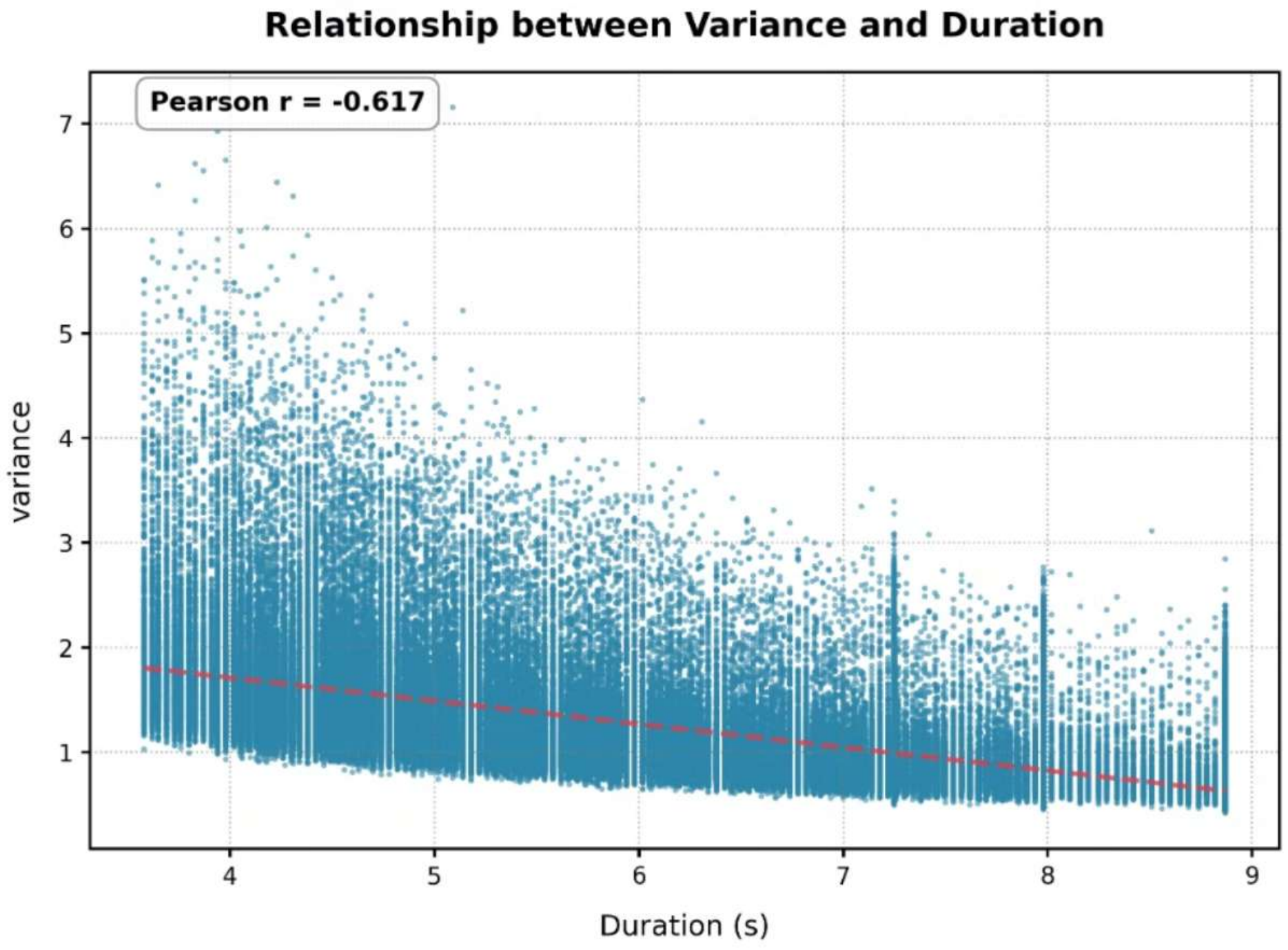}
    }
    \hfill
    \hspace{-6mm}
    \subfloat[$\mathcal{U}^3$-xi]{%
        \includegraphics[width=0.49\linewidth]{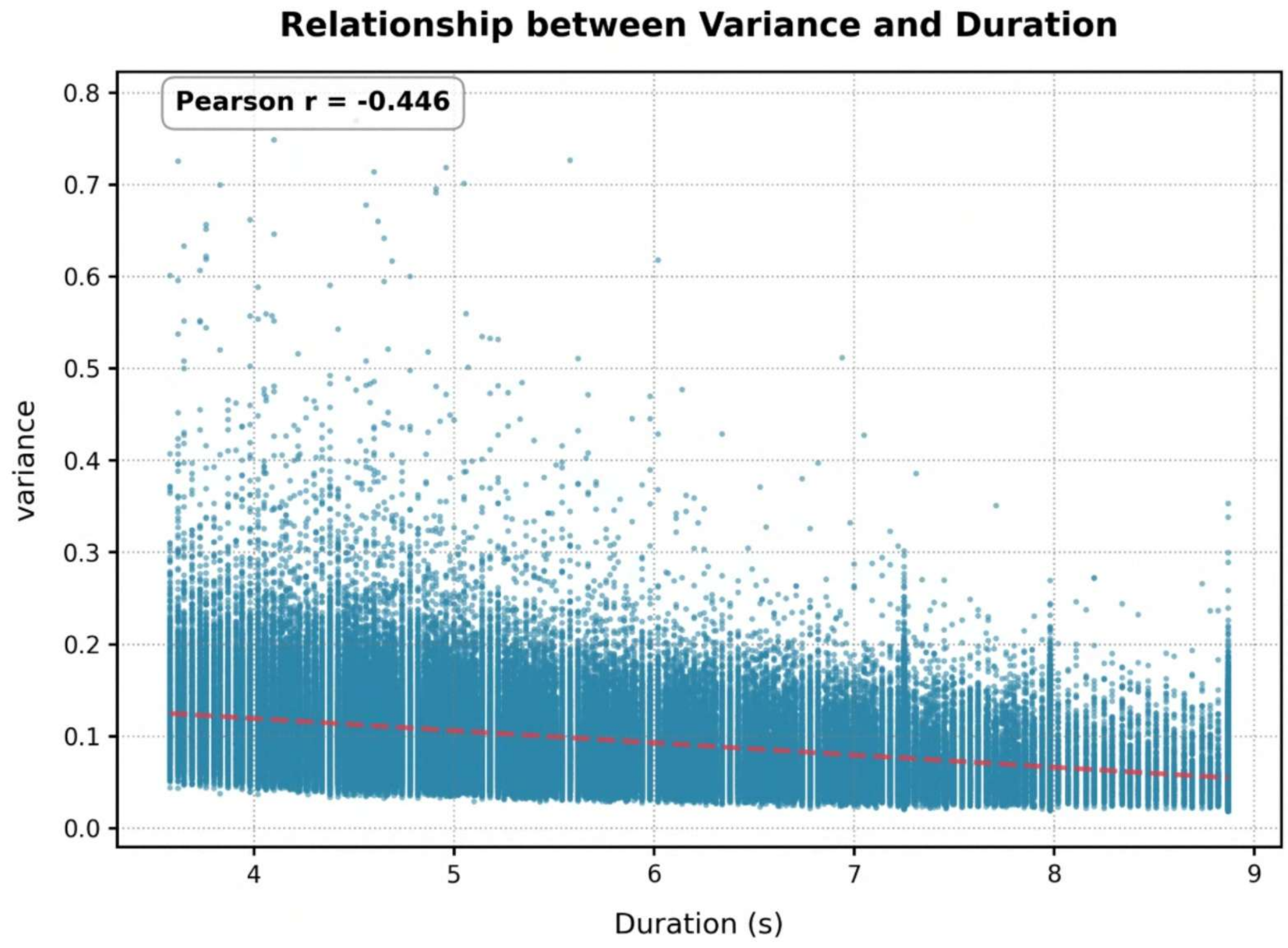}
    }

    \caption{
    Relationship between utterance-level variance (average variance over embedding dimension) and its duration on the Vox1 test set, including Vox1-O, Vox1-E and Vox1-H. 
    }\vspace{-3mm}
    \label{fig:variance_duration}
\end{figure}
\subsubsection{Relationship between Uncertainty and Sample Difficulty}
We first investigate the relationship between utterance-level uncertainty (i.e., the diagonal covariance $\mathbf{\Sigma}^\text{s}$) and the cosine gap
$\Delta_{\text{cos}}$,
which serves as a measure of the classification difficulty of sample $i$, where a larger gap indicates that the sample is easier to classify. 
Fig.~\ref{fig:variance_softmax} compares this relationship before and after applying our proposed method $\mathcal{U}^3$-xi. The Pearson correlation coefficient decreases from -0.166 (xi-vector) to -0.511 ($\mathcal{U}^3$-xi), demonstrating that the proposed method strengthens the negative correlation between uncertainty and sample hardness. This trend indicates that $\mathcal{U}^3$-xi produces more discriminative and interpretable uncertainty, assigning lower variance to easier samples and higher variance to harder ones. This demonstrates that our global-level uncertainty supervision indeed provides meaningful and effective guidance.

\begin{figure*}[htbp]
\vspace{-3mm}
    \centering
    \includegraphics[width=1\linewidth]{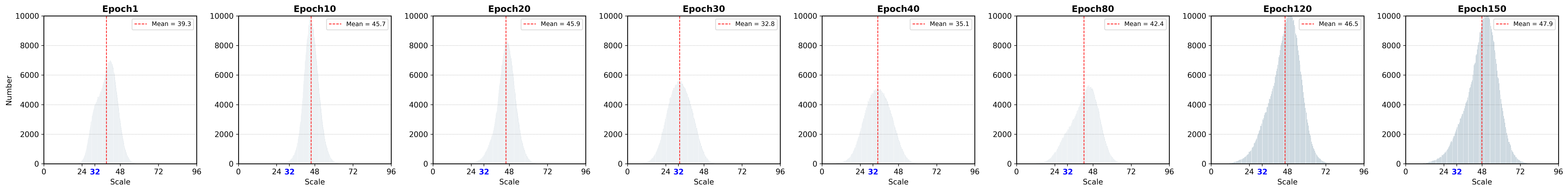}
    \vspace{-8mm}
    \caption{Distribution of the sample-wise scale across training epochs. 
    The x-axis represents the scale values, and the y-axis represents the number of samples corresponding to each scale value. The blue number 32 is the default scale. }
    \label{fig:scale_epoch}
    \vspace{-3mm}
\end{figure*}

\subsubsection{Relationship between Uncertainty and Sample Duration}
Fig.~\ref{fig:variance_duration} illustrates the relationship between utterance-level uncertainty and utterance duration on the test set.
Both subfigures show a clear trend that longer utterances tend to have lower estimated variance, which aligns with the intuitive understanding that embeddings extracted from longer speech segments are generally more robust. Interestingly, we observe that this correlation appears weaker for our proposed $\mathcal{U}^3$-xi model. Despite this weaker dependency on utterance duration, $\mathcal{U}^3$-xi still outperforms the baseline xi-vector model. One possible explanation is that $\mathcal{U}^3$-xi may capture aspects of utterance quality beyond length, allowing it to assign lower uncertainty even to short but high-quality utterances.

\subsubsection{The Distribution of Sample-wise Scale Across Training Epochs}

Fig.~\ref{fig:scale} suggests that a moderately larger scale can benefit
optimization. To further analyze this behavior, we visualize the evolution of
sample-wise scales across training epochs.
Since each utterance is assigned a single, sample-dependent scale, we summarize
the scale distribution at each epoch in Fig.~\ref{fig:scale_epoch}.
Several observations can be made:
\begin{enumerate}[label=\arabic*.]
\item \textbf{Sample-wise perspective:}
Within each epoch, different samples are assigned different scales.
Samples with large scales are typically easy and reliably classified,
whereas samples with small scales correspond to hard or unreliable cases.
This demonstrates that the model adaptively assigns sample-dependent scales
according to the difficulty of each utterance.

\item \textbf{Epoch-wise perspective:}
Before the angular margin starts increasing (epochs $\leq 20$), the mean scale
gradually rises.
When the margin begins to increase ($20 < \text{epoch} \leq 40$), the mean scale
drops noticeably, indicating the sensitivity of the scale to changes in the
margin constraint.
After the margin is fixed (epoch $> 40$), the mean scale increases again as
training proceeds.

This trend can be explained by the training dynamics.
As training progresses, an increasing number of samples become correctly
classified, resulting in a higher proportion of easy samples and thus a larger
mean scale at each epoch.
From a global optimization perspective, a larger average scale sharpens the softmax distribution, which relatively emphasizes misclassified or ambiguous samples by increasing their gradient contributions compared to previous training epochs.
\end{enumerate}
In summary, the dynamic scale effectively balances the contributions of training
utterances. The progressively right-shifting scale distribution further exhibits
a curriculum-learning behavior \cite{soviany2022curriculum}, where the model
initially emphasizes easier samples and gradually adapts to harder ones.

\begin{figure}[htbp]
\vspace{-3mm}
    \centering
    \includegraphics[width=0.9\linewidth]{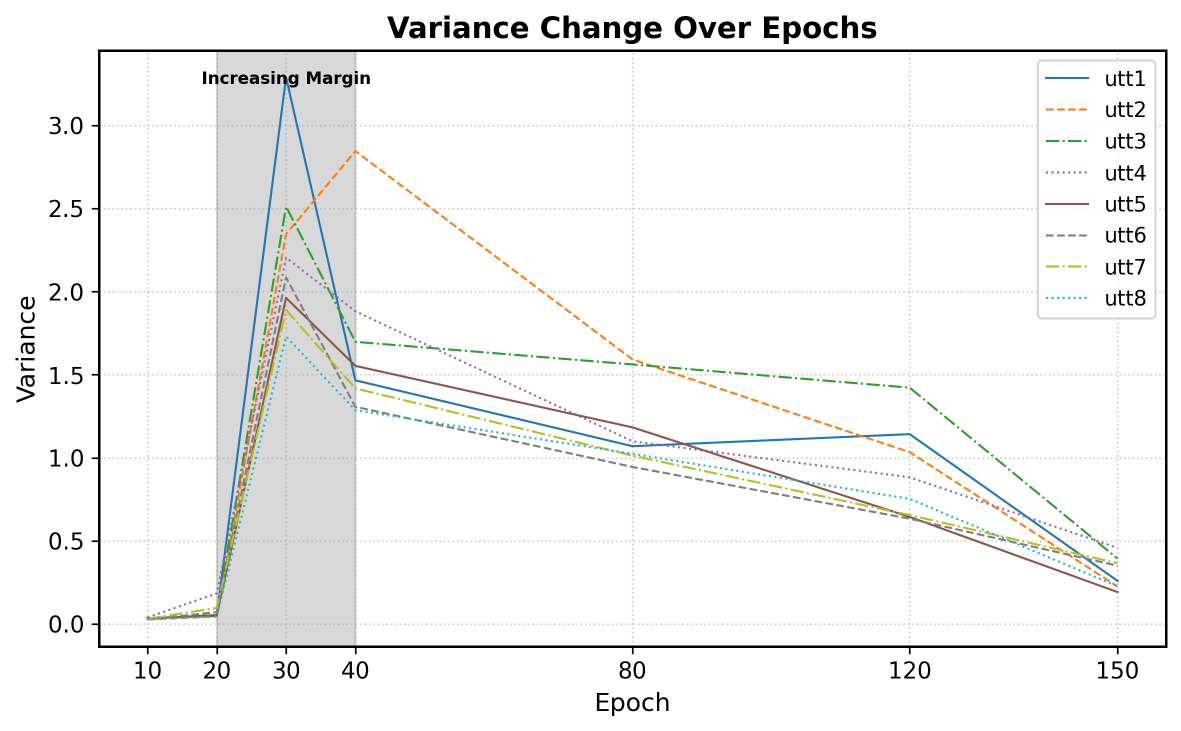}
    \vspace{-3mm}
    \caption{Uncertainty score (average variance over embedding dimension) of different utterances changes during training epochs. The margin is linearly increasing during epoch 20 to 40. The y-axis shows the averaged diagonal variance across all embedding dimensions}
    \label{fig:Uncertainty_change}
    \vspace{-3mm}
\end{figure}
\subsubsection{Uncertainty Dynamics during Training}

Fig.~\ref{fig:Uncertainty_change} illustrates the evolution of estimated
uncertainty across training epochs.
At early stages, uncertainty values remain low, reflecting the limited
discriminative capacity of the model.
As training proceeds, uncertainty gradually increases, and rises sharply when
the angular margin is introduced (epochs 20–40), indicating that the uncertainty
estimation module is highly sensitive to changes in the margin constraint.
After the margin is fixed, the uncertainty decreases steadily, suggesting that
the embedding space becomes more stable and the model converges toward more
confident representations.

Although our objective is to model data (aleatoric) uncertainty rather than model (epistemic) uncertainty, these observations suggest that the estimated data uncertainty is nevertheless influenced by the network’s optimization state. In particular, when the network has not yet converged to a stable solution, accurate estimation of data uncertainty becomes more challenging.

Furthermore, the relative uncertainty among different utterances evolves over
time.
For instance, an utterance that exhibits high uncertainty at intermediate
epochs may later become more confident than others.
This behavior suggests that uncertainty estimation is not reliable at early
training stages, but gradually adapts and becomes more meaningful as training
converges.
Overall, no consistent global ranking of uncertainty across utterances is
observed throughout training.

\renewcommand{\arraystretch}{0.9}
\begin{table*}[htp]
 
\setlength{\tabcolsep}{0.5pt} 

    \centering
\caption{Performance comparison across different speaker encoders. Cells highlighted in gray indicate the application of the proposed methods. `LM' denotes large-margin fine-tuning. `AS-Norm' denotes adaptive S-norm \cite{cumani11_interspeech}. `QMF' refers to the Quality Measure Function \cite{thienpondt2021idlab}. Results shown in gray correspond to the proposed methods. $\rho=1$ means uncertainty-aware scoring is applied. }
 
    \begin{tabular}{c l c c c c c c c|cc|c}
    \hline
    \toprule 
       \multirow{3}{*}{Row} & \multirow{2}{*}{Speaker \footnotemark} &\multirow{2}{*}{\# Param.} &\multirow{3}{*}{LM} & \multirow{2}{*}{AS-} & Q & \multicolumn{3}{c|}{In-domain Test} & \multicolumn{2}{c|}{Cross-domain Test} & Avg. $\Delta$  \\ \cline{7-11}
       &  & & & & M& Vox1-O & Vox1-E & Vox1-H  & SITW-Eval & CNCeleb& \multirow{2}{*}{In \textbar{} Cross}    \\ \cline{7-11}
       & Encoder&(M)   & & Norm& F& EER / minDCF & EER / minDCF & EER / minDCF & EER / minDCF & EER / minDCF &  \\ \hline \hline
1      & ECAPA-TDNN1024\cite{liu2025adaspeaker} & 14.65 &  \ding{56} &\ding{52} & \ding{52} & 0.660 / 0.078 & 0.882 / 0.094 & 1.606 / 0.159 & - & - & - \\ \hline
2      & ConFusionformer-9 \cite{tu2025confusionformer}& 10.9 &  \ding{56} &\ding{52} & \ding{56} & 0.68 / 0.064 &0.93 / 0.104 & 1.66 / 0.166 & - & - & - \\ \hline
3      & CAM++$^\dagger$ \cite{wang2023cam++} & 7.2 & \ding{56} &\ding{56} & \ding{56} & 0.808 / 0.109 & 0.931 /  0.109 & 1.863 / 0.179 & 1.704 / 0.166 &  15.179 / 0.635 & - \\ \hline
4      & Gemini SD-ResNet38 \cite{liu2024golden} & 6.72 & \ding{56} &\ding{56} & \ding{56} & 1.085 / 0.099 & 1.130 / 0.117 & 1.974 / 0.185 &1.523 / 0.147 & 11.507 / 0.553 & - \\ \hline 
5      & Gemini DF-ResNet114$^\dagger$ \cite{liu2024golden} &6.53 & \ding{52} &\ding{56} & \ding{56} & 0.771 / 0.065 & 0.906 / 0.097 & 1.599 / 0.156 &1.603 / 0.121  & 10.960 / 0.504  & -\\ \hline
6      & RecXi with L$_\text{ssp}$ \cite{liu2023disentangling} & 7.06 & \ding{56} &\ding{56} & \ding{56} & 0.984 / 0.091 & 1.075 / 0.114 & 1.857 / 0.179 & 1.340 / 0.137 & - &- \\ \hline
7      & ECAPA-TDNN1024 \cite{desplanques20_interspeech}$^\dagger$ & 14.65 &\ding{56} &\ding{56} & \ding{56}& 0.856 / 0.090 & 1.072 / 0.117 & 2.059 / 0.205 &2.269 / 0.174 & 15.532 / 0.670 & - \\ \hline 
8      & \multirow{1}{*}{ ECAPA-TDNN512 \cite{desplanques20_interspeech}}$^\dagger$& 6.19 &\ding{56} &\ding{56} & \ding{56}& 1.069 / 0.122 & 1.209 / 0.136 & 2.310 / 0.226 & 2.433 / 0.197 & 15.314 / 0.633 & Benchmark \\ \cline{3-12}  \hhline{|~~~---|} 
9      & \quad+ $\mathcal{U}^3$-xi ($\rho=0$) &  \multirow{2}{*}{6.69} & \ding{56} &\ding{56} & \ding{56} & \cellcolor{gray!20} 0.856 / 0.109 & \cellcolor{gray!20}1.064 / 0.121 & \cellcolor{gray!20}1.982 / 0.195 & \cellcolor{gray!20}2.105 / 0.170 &  \cellcolor{gray!20}13.706 / 0.608 & \cellcolor{gray!20}13.57\% \textbar{} 10.41\% \\ 
10     & \quad+ $\mathcal{U}^3$-xi ($\rho=1$) & & \ding{56} &\ding{56} & \ding{56} & \cellcolor{gray!20} 0.782 / 0.100 & \cellcolor{gray!20}1.016 / 0.115 & \cellcolor{gray!20}1.888 / 0.187 & \cellcolor{gray!20}1.921 / 0.169 &  \cellcolor{gray!20}10.271 / 1.000 &\cellcolor{gray!20} 18.64\% \textbar{} 2.54\%  \\  \hline
11     & \multirow{1}{*}{ResNet34 \cite{zeinali2019but}}$^\dagger$ &6.63 &\ding{56} &\ding{56} & \ding{56} & 0.867 / 0.091 & 1.049 / 0.121 & 1.960 / 0.192 & 1.903 / 0.159 & 11.090 / \textbf{0.488} & Benchmark \\ \cline{3-12}  \hhline{|~~~---|} 
12     & \quad + $\mathcal{U}^3$-xi ($\rho=0$)& \multirow{2}{*}{7.92} &\ding{56} &\ding{56} &\ding{56}&  \cellcolor{gray!20}0.888 / 0.085 &\cellcolor{gray!20} 0.900 / 0.099 & \cellcolor{gray!20}1.712 / 0.175 & \cellcolor{gray!20}1.435 / 0.135 & \cellcolor{gray!20}11.732 / 0.513 &  \cellcolor{gray!20}9.59\% \textbar{} 7.19\%\\
13     & \quad+ $\mathcal{U}^3$-xi ($\rho=1$) &  &\ding{56} &\ding{56} &\ding{56}&\cellcolor{gray!20}0.867 / 0.078 &\cellcolor{gray!20} 0.868 / 0.095 &\cellcolor{gray!20} 1.641 / 0.172 &\cellcolor{gray!20} 1.367 / 0.134 &\cellcolor{gray!20}10.082 / 0.541&\cellcolor{gray!20} 13.29\% \textbar{} 10.53\%  \\  \hline 
14     & \multirow{4}{*}{ReDimNet-B2$^\ddagger$ \cite{yakovlev24_interspeech}} & \multirow{4}{*}{4.89} & \ding{56} &\ding{56} & \ding{56}  & 0.782 / 0.064 & 0.907 / 0.097 & 1.667 / 0.162 & 1.558 / 0.148 & 12.385 / 0.552 & Benchmark\\ 
15     &  &  &  \ding{52} &\ding{56} & \ding{56} & 0.750 / 0.072 & 0.868 / 0.093 & 1.560 / 0.147 & 1.293 / 0.120 & 12.509 / 0.565 & 2.47\% \textbar{} 8.14\%   \\ 
16     &  &  &  \ding{52} &\ding{52} & \ding{56} & 0.675 / 0.063 & 0.826 / 0.088 & 1.437 / 0.135 & 1.213 / 0.154 & 10.662 / 0.518 & 10.66\% \textbar{} 9.55\%   \\ 

17   &  &  &  \ding{52} &\ding{52} & \ding{52} & 0.686 / \textbf{0.060} & 0.805 / 0.084 & 1.395 / 0.130 & 1.176 / 0.155 & 9.254 / 0.585 &  13.21\% \textbar{} 9.77\% \\ \cline{3-12} \hhline{|~~~~~~------|} 

18     & \quad + $\mathcal{U}^3$-xi ($\rho=0$) &  & \multirow{2}{*}{\ding{56}} & \multirow{2}{*}{\ding{56}} & \multirow{2}{*}{\ding{56}} & \cellcolor{gray!20}0.649 / \cellcolor{gray!20}0.073 & \cellcolor{gray!20}0.801 / \cellcolor{gray!20}0.089 & \cellcolor{gray!20}1.532 / \cellcolor{gray!20}0.153 & \cellcolor{gray!20}1.422 / \cellcolor{gray!20}0.127 &  \cellcolor{gray!20}13.464 / \cellcolor{gray!20}0.552 & \cellcolor{gray!20}5.76\% \textbar{} \cellcolor{gray!20}3.55\% \\ 
19     & \quad + $\mathcal{U}^3$-xi ($\rho=1$)&   & & &    & \cellcolor{gray!20}0.606 / \cellcolor{gray!20}0.065 & \cellcolor{gray!20}0.779 / \cellcolor{gray!20}0.091 & \cellcolor{gray!20}1.494 / \cellcolor{gray!20}0.157 & \cellcolor{gray!20}1.394 / \cellcolor{gray!20}0.141 & \cellcolor{gray!20}\textbf{9.479} / \cellcolor{gray!20}1.000 & \cellcolor{gray!20}9.45\% \textbar{} \cellcolor{gray!20}-10.63\%  \\  \hhline{|~~~~~~------|} 
20     & \quad + $\mathcal{U}^3$-xi ($\rho=0$)&   &   \multirow{2}{*}{\ding{52}} & \multirow{2}{*}{\ding{56}} & \multirow{2}{*}{\ding{56}} & \cellcolor{gray!20}0.670 / \cellcolor{gray!20}0.070 & \cellcolor{gray!20}0.770 / \cellcolor{gray!20}0.081 & \cellcolor{gray!20}1.437 / \cellcolor{gray!20}0.144 &  \cellcolor{gray!20}1.230 / \cellcolor{gray!20}\textbf{0.098} & \cellcolor{gray!20}12.877 / \cellcolor{gray!20}0.622 & \cellcolor{gray!20}10.88\% \textbar{} \cellcolor{gray!20}9.04\%  \\
21     & \quad + $\mathcal{U}^3$-xi ($\rho=1$) & 5.46 & & & &   \cellcolor{gray!20}0.489 / \cellcolor{gray!20}0.076 & \cellcolor{gray!20}0.698 / \cellcolor{gray!20}0.104 & \cellcolor{gray!20}1.311 / \cellcolor{gray!20}0.203 & \cellcolor{gray!20}1.017 / \cellcolor{gray!20}0.211 &  \cellcolor{gray!20}10.267 / \cellcolor{gray!20}1.000 & \cellcolor{gray!20}1.60\% \textbar{} \cellcolor{gray!20}-13.45\%  \\ \hhline{|~~~~~~------|} 
22     & \quad + $\mathcal{U}^3$-xi ($\rho=0$) & & \multirow{2}{*}{\ding{52}} & \multirow{2}{*}{\ding{52}} & \multirow{2}{*}{\ding{56}} &  \cellcolor{gray!20}0.622 / \cellcolor{gray!20}0.069 & \cellcolor{gray!20}0.726 / \cellcolor{gray!20}0.074 & \cellcolor{gray!20}1.327 / \cellcolor{gray!20}0.126 & \cellcolor{gray!20}1.093 / \cellcolor{gray!20}0.125 & \cellcolor{gray!20}11.462 / \cellcolor{gray!20}0.740 & \cellcolor{gray!20}16.47\% \textbar{} \cellcolor{gray!20}4.70\% \\
23     & \quad + $\mathcal{U}^3$-xi ($\rho=1$)&  &   &&  & \cellcolor{gray!20}\textbf{0.399} / \cellcolor{gray!20}0.080 & \cellcolor{gray!20}\textbf{0.638} / \cellcolor{gray!20}0.089 & \cellcolor{gray!20}\textbf{1.170} / \cellcolor{gray!20}0.171 & \cellcolor{gray!20}\textbf{0.875} / \cellcolor{gray!20}0.224 & \cellcolor{gray!20}9.738 / \cellcolor{gray!20}1.000& \cellcolor{gray!20}16.22\% \textbar{} \cellcolor{gray!20}-16.58\%   \\   \hhline{|~~~~~~------|} 
24     & \quad + $\mathcal{U}^3$-xi ($\rho=0$) & & \multirow{2}{*}{\ding{52}} & \multirow{2}{*}{\ding{52}} & \multirow{2}{*}{\ding{52}} & \cellcolor{gray!20}0.596 / \cellcolor{gray!20}0.070 & \cellcolor{gray!20}0.705 / \cellcolor{gray!20}0.071 & \cellcolor{gray!20}1.298 / \cellcolor{gray!20}0.124 & \cellcolor{gray!20} \textbf{0.875} / 0.227 &\cellcolor{gray!20} 10.115 / 0.832 & \cellcolor{gray!20}18.18\% \textbar{} -10.48\% \\ 
25 & \quad + $\mathcal{U}^3$-xi ($\rho=1$)  & & &   &  & \cellcolor{gray!20}0.569 / \cellcolor{gray!20}0.064 & \cellcolor{gray!20}0.679 / \cellcolor{gray!20}\textbf{0.068} & \cellcolor{gray!20}1.234 / \cellcolor{gray!20}\textbf{0.121} & \cellcolor{gray!20}1.066 / 0.131 &\cellcolor{gray!20} 10.436 / 1.000 & \cellcolor{gray!20}22.26\% \textbar{} -5.59\% \\

      \hline   
        
         \bottomrule
    \end{tabular}
    
    \label{tab:encoder}
    \vspace{-3mm}
\end{table*}
\footnotetext{$^\dagger$ indicates that these results are directly inferred using the pretrained WeSpeaker model available at \url{https://github.com/wenet-e2e/wespeaker/blob/master/docs/pretrained.md}. \
$^\ddagger$ means we retrain these models by ourselves. }

\subsection{Generalization Ability to Other Speaker Encoders}
Table \ref{tab:encoder} presents a comparison between our methods and other state-of-the-art approaches. The results demonstrate that our method achieves competitive, and in some cases superior, performance while using fewer parameters. 
\subsubsection{In-domain Test} 
Considering ECAPA-TDNN512 (Rows 8–10), ResNet34 (Rows 11–13), and ReDimNet-B2 (Rows 14, 18, and 19), we observe that the proposed $\mathcal{U}^3$-xi consistently achieves performance improvements over their corresponding baselines. Furthermore, applying uncertainty-aware scoring with $\rho = 1$ generally leads to additional gains. The only exceptions are ReDimNet-B2 evaluated on Vox1-E and Vox1-H, where minDCF slightly increases from 0.089 to 0.091 and from 0.153 to 0.157, respectively.
In terms of average performance, the relative improvements are 18.64\% for ECAPA-TDNN512, 13.29\% for ResNet34, and 9.45\% for ReDimNet-B2. These results indicate that while $\mathcal{U}^3$-xi remains effective across different speaker encoders, the improvement diminishes as the speaker encoder becomes stronger.

\subsubsection{Cross-domain Test}
The SITW-EVAL set is relatively similar to VoxCeleb in terms of recording conditions; consequently, its performance trends are consistent with those observed in the in-domain evaluation. Specifically, setting $\rho=0$ yields consistent improvements across different speaker encoders in terms of all metrics, whereas $\rho=1$ results in a slight degradation in terms of minDCF.

In contrast, on the more challenging cross-domain CNCeleb dataset, setting $\rho=0$ fails to achieve stable improvements in either EER or minDCF.
Moreover, setting $\rho=1$ consistently improves EER; however, it leads to further degradation in terms of minDCF in several cases, with severe failures observed in Rows 10 and 19, where the minDCF spikes to 1.0, clearly indicating invalid operating points.

\begin{figure}[htbp]
    \centering
\vspace{-3mm}
    \subfloat[Row 9]{%
        \includegraphics[width=0.49\linewidth]{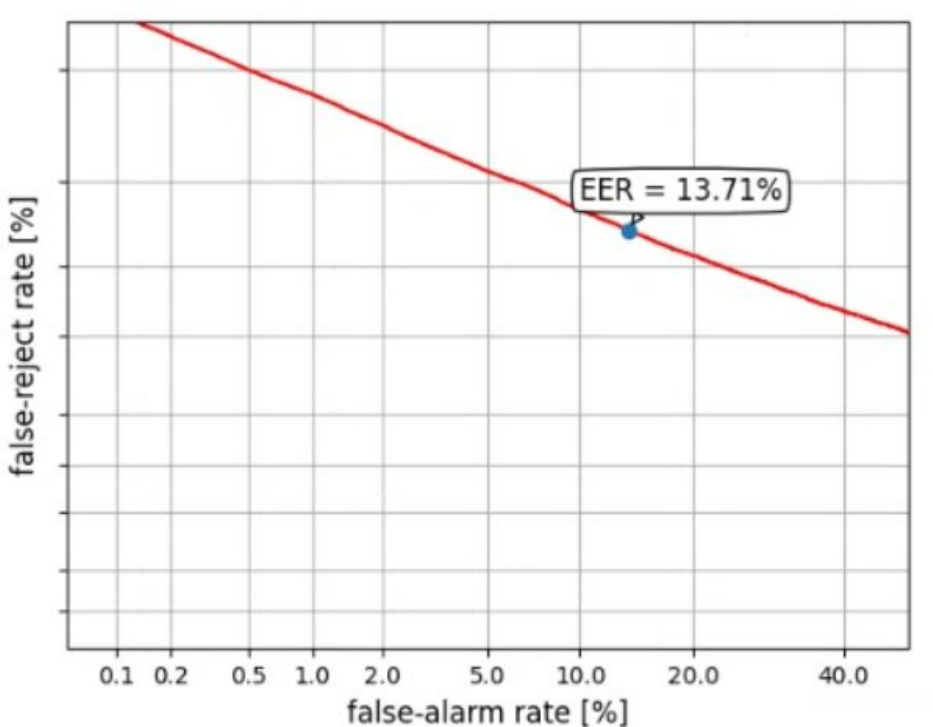}
    }
    \hfill
    \hspace{-2mm}
    \subfloat[Row 10 ]{%
        \includegraphics[width=0.49\linewidth]{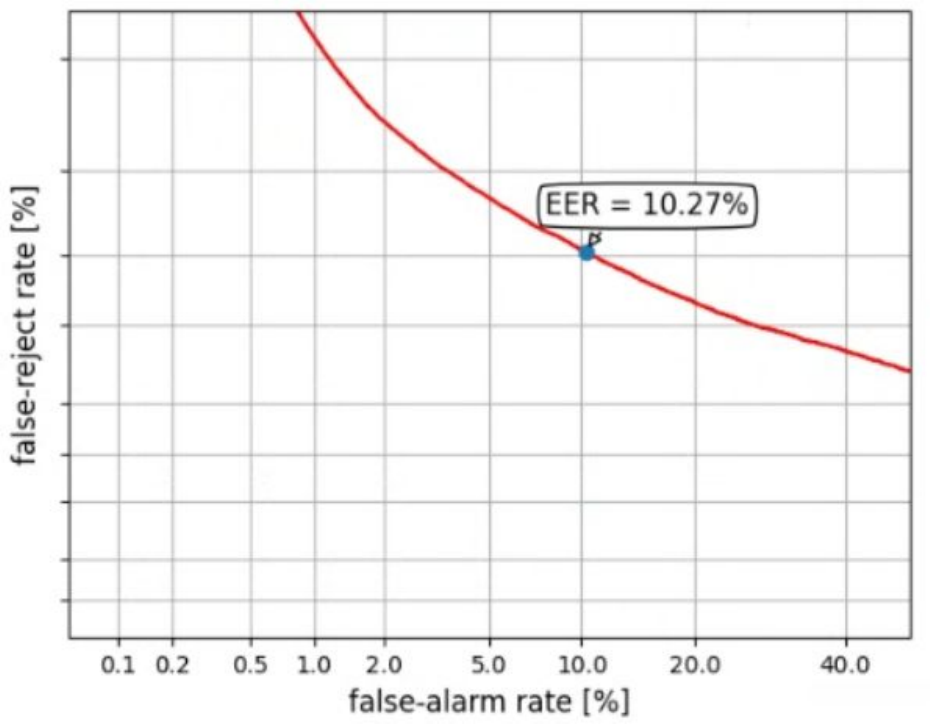}
    }

    \caption{The Detection Error Tradeoff (DET) curve on CNCeleb dataset. 
    }
    \label{fig:epcapa_cncelb}
    \vspace{-3mm}
\end{figure}

To further investigate this issue, we plot the Detection Error Tradeoff (DET) curves for Rows 9 and 10, as shown in Fig. \ref{fig:epcapa_cncelb}. It can be observed that the DET curve of Row 10 becomes severely distorted. Compared with the relatively smooth and well-balanced curve of Row 9, the curve of Row 10 exhibits a much steeper and less stable tradeoff between the false-alarm rate (FAR) and the false-reject rate (FRR).

This distortion suggests that uncertainty-aware scoring, while optimizing EER (i.e., the operating point where FAR and FRR are equal), fails to preserve a stable tradeoff across a broader range of operating points. As a result, this imbalance directly leads to erratic minDCF behavior, including the extreme values observed on CNCeleb.

We hypothesize that this limitation arises from poor generalization of the uncertainty estimation mechanism to cross-domain conditions. During training on VoxCeleb2, variations in recording conditions may be implicitly modeled as a form of uncertainty. However, in cross-domain scenarios such as CNCeleb, the acoustic conditions differ substantially from those seen during training, causing the uncertainty estimator to fail in producing accurate or meaningful uncertainty estimates.



\subsubsection{Effect of LM, AS-Norm, and QMF on Uncertainty Modeling}

We further investigate the compatibility of the proposed $\mathcal{U}^3$-xi
framework with commonly used training-time fine-tuning and backend score
normalization techniques, including large-margin (LM) fine-tuning, AS-Norm,
and QMF (Rows 15--17 and 20--25 in Table~\ref{tab:encoder}).
Overall, these techniques remain compatible with $\mathcal{U}^3$-xi and can
provide additional gains; however, their interaction with uncertainty-aware
scoring exhibits distinct behaviors across evaluation settings.

\begin{enumerate}[label=\arabic*.]

\item 
Under the LM setting (Rows 15, 20, and 21), fixing $\rho=0$ consistently improves
performance over the LM baseline (Row 15) on most evaluation sets, except for
the cross-domain CNCeleb dataset.
When enabling uncertainty-aware scoring with $\rho=1$, consistent improvements
are observed in terms of EER across all test sets, but at the cost of degraded
minDCF.
We attribute this behavior to the high sensitivity of uncertainty estimation
to changes in the angular margin.
In particular, the large margin is applied for only a small number of epochs
(five epochs in total), which may be insufficient for the uncertainty estimation
module to converge to a stable and well-calibrated solution.

\item 
A similar pattern is observed when AS-Norm is applied (Rows 16, 22, and 23).
Setting $\rho=0$ yields consistent improvements over the AS-Norm baseline
(Row 16) on most evaluation sets, except for CNCeleb.
In contrast, setting $\rho=1$ consistently improves EER while degrading minDCF
across all test sets, indicating that AS-Norm does not fully resolve the
calibration instability introduced by uncertainty-aware scoring.

\item 
QMF partially alleviates the above issues, as shown by the comparison among
Rows 17, 24, and 25.
For in-domain evaluations, both Rows 24 and 25 consistently outperform the QMF
baseline (Row 17) on most test sets in terms of either EER or minDCF, even when
uncertainty-aware scoring with $\rho=1$ is enabled.
However, on cross-domain data, the performance gains remain unstable.
Specifically, improvements are only observed on SITW-Eval, with Row 24 improving
EER and Row 25 improving minDCF.
This suggests that while QMF enhances robustness to uncertainty-aware scoring,
domain mismatch remains a fundamental challenge for reliable uncertainty
calibration.
\end{enumerate}

\vspace{-3mm}
\section{Conclusion and Limitations}
We have studied uncertainty modeling in speaker recognition and propose two uncertainty
supervision strategies together with a redesigned estimation module.
These components are unified in the proposed $\mathcal{U}^3$-xi framework,
which yields consistent and substantial performance gains across diverse
experimental settings.
The results demonstrate that explicitly modeling uncertainty is an effective
and promising direction for improving speaker recognition. 
Despite these gains, uncertainty-aware scoring does not consistently improve
minDCF, particularly under cross-domain evaluation.
This suggests that uncertainty estimation remains sensitive to domain mismatch
and may not yet be sufficiently calibrated in such conditions.
Future work will focus on more robust uncertainty modeling and on scaling the
framework to larger and more diverse training corpora.





%
\section{ACKNOWLEDGMENTS}
We would like to thank Jinghan Liu, Tianchi Liu, and Yi Ma for their kind assistance, which has provided substantial inspiration for completing this paper.

\bibliographystyle{IEEEtran}
\bibliography{IEEEabrv,Bibliography}

\vfill

\newpage

\end{document}